\documentclass[useAMS,usenatbib]{mn2e}
\usepackage{psfig, epsf, epsfig}

\title[Thick outflow in the M\,101 ULS]{Revisiting the ultraluminous 
supersoft source in M\,101: \\
an optically thick outflow model}
\author[R. Soria \& A.K.H. Kong]{Roberto Soria$^{1}$\thanks{E-mail:
roberto.soria@icrar.org (RS); akong@phys.nthu.edu.tw (AKHK)}, 
Albert K.~H.~Kong$^{2}$\\
$^{1}$International Centre for Radio Astronomy Research, Curtin University, 
GPO Box U1987, Perth, WA 6845, Australia\\
$^{2}$Institute of Astronomy and Department of Physics, National Tsing Hua University, Hsinchu 30013, Taiwan}
\begin{document}

\date{Accepted 2015 November 11}

\pagerange{\pageref{firstpage}--\pageref{lastpage}} \pubyear{2011}

\maketitle

\label{firstpage}

\begin{abstract}
The M\,101 galaxy contains the best-known example of an ultraluminous supersoft 
source (ULS), dominated by a thermal component at $kT \approx 0.1$ keV. 
The origin of the thermal component and the relation between ULSs 
and standard (broad-band spectrum) ultraluminous X-ray sources (ULXs) 
are still controversial. We re-examined the X-ray spectral 
and timing properties of the M\,101 ULS using archival 
{\it Chandra} and {\it XMM-Newton} observations. 
We show that the X-ray time-variability and spectral properties 
are inconsistent with standard disk emission. 
The characteristic radius $R_{\rm bb}$ of the thermal 
emitter varies from epoch to epoch between $\approx$10,000 km 
and $\approx$100,000 km; the colour temperature $kT_{\rm bb}$ varies 
between $\approx$50 eV and $\approx$140 eV; and the two quantities scale 
approximately as $R_{\rm bb} \propto T_{\rm bb}^{-2}$. 
In addition to the smooth continuum, we also find (at some 
epochs) spectral residuals well fitted with thermal 
plasma models and absorption edges: we interpret this as evidence 
that we are looking at a clumpy, multi-temperature outflow. 
We suggest that at sufficiently high accretion rates and inclination angles, 
the super-critical, radiatively driven outflow becomes effectively optically 
thick and completely thermalizes the harder X-ray photons from the inner 
part of the inflow, removing the hard spectral tail. 
We develop a simple, spherically symmetric outflow model and 
show that it is consistent with the observed temperatures, 
radii and luminosities. 
A larger, cooler photosphere 
shifts the emission peak into the far-UV and makes the source dimmer 
in X-rays but possibly ultraluminous in the UV.
We compare our results and interpretation with those of \cite{liu13}.
  
\end{abstract}

\begin{keywords}
accretion, accretion discs -- X-rays: individual: M\,101 ULX-1 -- black hole physics.
\end{keywords}

\section{Introduction}

Ultraluminous X-ray sources (ULXs) are the highest-luminosity group 
of the X-ray binary population, empirically defined by an X-ray luminosity 
$L_{\rm X} \ga 3 \times 10^{39}$ erg s$^{-1}$  (\citealt{feng11} for a review).
In most cases, the simplest explanation consistent with the observations 
is that ULXs contain a stellar-mass black hole (BH) accreting above 
its Eddington limit \citep{gladstone09,sutton13,motch14}. 
In our Galaxy, the masses of stellar BHs are clustered around $\approx$5--15$M_{\odot}$ \citep{kreidberg12}; however, in lower-metallicity galaxies, stellar evolution models allow for the formation of BHs as massive as $\approx$80$M_{\odot}$ \citep{belczynski10}, corresponding to an Eddington luminosity $\approx 10^{40}$ erg s$^{-1}$.
When the accretion rate is super-critical 
($\dot{m} \equiv 0.1 \dot{M}c^2 / L_{\rm Edd} > 1$), 
the photon luminosity mildly exceeds the Eddington limit: 
$L \approx L_{\rm Edd} (1 + a \ln \dot{m})$ 
where $3/5 \la a \la 1$ depending on the relative fraction of energy 
carried by outflows or advected through the BH horizon 
\citep{poutanen07,ss73}.

\begin{table*}
\begin{center}
\begin{tabular}{lcccr}\hline
\hline
Date & ObsID & Exposure time & $0.3$--2 keV Net Count Rate & $L^{\rm min}_{0.3-10}$\\[5pt]
 & & (ks) & ($10^{-4}$ ct s$^{-1}$) & (erg s$^{-1}$)\\
\hline\\[-5pt]
2000 Mar 26 & 934 & 98.38 & $914 \pm 10$ & $\left(1.6^{+0.1}_{-0.1}\right) \times 10^{39}$\\[5pt]
2000 Oct 29 & 2065 & 9.63 & $297 \pm 18$ & $\left(6.3^{+0.5}_{-0.9}\right) \times 10^{38}$ \\[5pt]
2004 Jan 19 & 4731 & 56.24 & $3.0 \pm 0.8$ & $\left(1.7^{+0.5}_{-0.5}\right) \times 10^{37}$ \\[5pt]
2004 Jan 24 & 5297 & 21.69 & $4.7 \pm 1.5$ & $\left(2.7^{+0.8}_{-0.8}\right) \times 10^{37}$ \\[5pt]
2004 Mar 07 & 5300 & 52.09 & $4.3 \pm 1.0$ & $\left(2.5^{+0.6}_{-0.6}\right) \times 10^{37}$  \\[5pt]
2004 Mar 14 & 5309 & 70.77 & $2.3 \pm 0.6$ & $\left(1.3^{+0.3}_{-0.3}\right) \times 10^{37}$ \\[5pt]
2004 Mar 19 & 4732 & 69.79 & $2.0 \pm 0.6$ & $\left(1.2^{+0.4}_{-0.4}\right) \times 10^{37}$ \\[5pt]
2004 May 03 & 5322 & 64.7 & $1.8 \pm 0.6$ & $\left(1.0^{+0.3}_{-0.3}\right) \times 10^{37}$ \\[5pt]
2004 May 07 & 4733 & 24.81 & $2.2 \pm 1.1$ & $\left(1.3^{+0.7}_{-0.7}\right) \times 10^{37}$\\[5pt]
2004 May 09 & 5323 & 42.62 & $1.9 \pm 0.7$ & $\left(1.1^{+0.4}_{-0.4}\right) \times 10^{37}$\\[5pt]
2004 Jul 05 & 5337 & 9.94 & $196 \pm 14$ & $\left(9.0^{+1.0}_{-1.0}\right) \times 10^{38}$ \\[5pt]
2004 Jul 06 & 5338 & 28.57 & $260 \pm 10$ & $\left(9.4^{+0.8}_{-1.0}\right) \times 10^{38}$ \\[5pt]
2004 Jul 07 & 5339 & 14.32 & $231 \pm 13$ & $\left(8.4^{+1.0}_{-1.0}\right) \times 10^{38}$\\[5pt]
2004 Jul 08 & 5340 & 54.42 & $102 \pm 4$ & $\left(4.5^{+1.0}_{-0.6}\right) \times 10^{38}$\\[5pt]
2004 Jul 11 & 4734 & 35.48 & $67 \pm 4$ & $\left(4.0^{+0.6}_{-0.6}\right) \times 10^{38}$\\[5pt]
2004 Jul 23 & 0164560701 & 21.1 & $58 \pm 8$ & $\left(6.8^{+1.3}_{-1.3}\right) \times 10^{37}$\\[5pt]
2004 Sep 05 & 6114 & 66.2 & $3.0 \pm 0.7$ & $\left(1.7^{+0.4}_{-0.4}\right) \times 10^{37}$\\[5pt]
2004 Sep 08 & 6115 & 35.76 & $2.5 \pm 0.9$ & $\left(1.4^{+0.6}_{-0.6}\right) \times 10^{37}$ \\[5pt]
2004 Sep 11 & 6118 & 11.46 & $3.9 \pm 2.0$ & $\left(2.3^{+1.2}_{-1.2}\right) \times 10^{37}$\\[5pt]
2004 Sep 12 & 4735 & 28.78 & $4.8 \pm 1.3$ & $\left(2.8^{+0.7}_{-0.7}\right) \times 10^{37}$\\[5pt]
2004 Nov 01 & 4736 & 77.35 & $2.3 \pm 0.6$ & $\left(1.3^{+0.3}_{-0.3}\right) \times 10^{37}$\\[5pt]
2004 Nov 07 & 6152 & 44.09 & $1.2 \pm 0.7$ & $\left(0.7^{+0.4}_{-0.4}\right) \times 10^{37}$\\[5pt]
2004 Dec 22 & 6170 & 47.95 & $8.3 \pm 1.3$ & $\left(4.1^{+0.7}_{-0.7}\right) \times 10^{37}$\\[5pt]
2004 Dec 24 & 6175 & 40.66 & $13.1 \pm 1.8$ & $\left(7.0^{+1.0}_{-1.0}\right) \times 10^{37}$\\[5pt]
2004 Dec 30 & 6169 & 29.38 & $210 \pm 8$ & $\left(7.4^{+0.7}_{-0.4}\right) \times 10^{38}$ \\[5pt]
2005 Jan 01 & 4737 & 21.85 & $664 \pm 17$ & $\left(1.6^{+0.1}_{-0.1}\right) \times 10^{39}$\\[5pt]
2005 Jan 08 & 0212480201 & 13.7 & $447 \pm 15$ & $\left(4.6^{+0.2}_{-0.2}\right) \times 10^{38}$\\
\hline
\end{tabular} 
\end{center}
\caption{Log of the observations considered for this study, and net count rate 
in the $0.3$--2 keV band. $L^{\rm min}_{0.3-10}$ 
is the inferred $0.3$--$10$ keV luminosity (defined as $4\pi d^2 \times$ observed flux) 
corrected {\it only} for the line-of-sight Galactic absorption 
$n_{\rm H} = 1.5 \times 10^{20}$ cm$^{-2}$; 
this is a strong, almost model-independent lower limit to the bolometric luminosity. 
Errors are 68\% confidence limits. All the entries in this table are 
{\it Chandra}/ACIS observations, except for 2004 July 23 and 2005 January 8, 
which are {\it XMM-Newton}/EPIC observations. The net count rate 
for those two observation is the combined rate of pn, MOS1 and MOS2.
}
\label{tab1}
\end{table*}

The X-ray spectral appearance of ULXs depends both on accretion rate 
and, for a given $\dot{m}$, on the viewing angle \citep{sutton13}. 
Sources that only mildly exceed the critical rate ($\dot{m} \la$ a few)
have a curved, thermal spectrum consistent with a 
non-standard accretion disk (slim disk models: 
\citealt{watarai01,mizuno01,kubota04}), 
with characteristic inner-disk temperatures $kT_{\rm in} \approx 1.3$--$2$ keV 
and a flatter radial temperature profile.
At higher accretion rates, the X-ray spectra of most ULXs show a slightly curved broad-band component dominating the 1--10 keV band, with an additional thermal component (soft excess) at $kT_{\rm bb} \approx 0.15$--0.30 keV. The origin of both components is still disputed. 
The broader component could come either from the inner part of the non-standard accretion disk, or from inverse-Compton scattering of the inner disk emission in a warm ($kT_e \sim 2$keV), optically thick corona \citep{middleton15,mukherjee15, walton15,roberts07}; the softer emission might come from a radiatively driven outflow, launched near or just outside the spherization radius \citep{king03,poutanen07,sutton13,middleton15}, or also from the disk \citep{miller13}.
The broad-band continuum becomes steeper, with a characteristic downturn at lower energies ($E \approx 5$ keV) for sources seen at higher inclination angles, probably because the X-ray photons in our line of sight pass through and are downscattered by a thicker disk wind \citep[soft ultraluminous regime:][]{sutton13}. Instead, ULXs seen at low inclination angles have a harder spectrum (hard ultraluminous regime), consistent with the interpretation that higher-energy photons emitted in the innermost part of the inflow can emerge from the low-density polar funnel with less downscattering. This scenario is consistent with numerical 
and theoretical models of massive radiatively-driven outflows in the super-critical regime \citep{poutanen07,dotan11,ohsuga11,kawashima12}.

Two rare subclasses of ULXs remain hard to explain with the super-Eddington 
stellar-mass BH scenario. The first case is that of ``hyperluminous'' 
X-ray sources, that is those very few ULXs 
(most notably, ESO\,243$-$49 HLX-1: \citealt{farrell09}, and 
M\,82 X41.4$+$60: \citealt{feng10,pasham14}) that reach 
$L_{\rm X} \ga 10^{41}$ erg s$^{-1}$, two orders of magnitude higher 
than the Eddington limit of an ordinary $10M_{\odot}$ BH.
For those sources, an intermediate-mass BH with $M > 100 M_{\odot}$ 
is the most likely explanation. 
The hyperluminous subclass is outside the scope of this work.

The second unexplained subclass, which we discuss in this paper, 
is that of ``supersoft'' ULXs \citep{distefano03}, henceforth referred 
to as ultraluminous supersoft sources (ULSs). 
A purely empirical definition of ULS is a source that is dominated by 
soft, blackbody-like emission with a hardness ratio (M$-$S)/T $\la -0.8$ 
(where S, M and T are the {\it Chandra} 
or {\it XMM-Newton} count rates in the 0.3--1.1 keV, 1.1--2.5 keV 
and 0.3--7.0 keV bands), and that has reached an extrapolated bolometric 
luminosity of the thermal component $L^{\rm bb}_{\rm bol} \ga 10^{39}$ erg s$^{-1}$ 
in at least one observation. 
The best-known representatives of this group have been detected in  
M\,101 \citep{kong04,kong05,mukai05,liu09,liu13},
M\,81 \citep{swartz02,liu08a,liu08b}, 
M\,51 \citep{distefano03,terashima04},
NGC\,300 \citep{read01,kong03}, 
NGC\,4631 \citep{vogler96,carpano07,soria09}, 
NGC\,247 \citep{jin11,tao12}
and the Antennae \citep{fabbiano03}.
A detailed study of the common properties of this whole group of ULSs 
is presented in a companion paper \citep{urquhart15}.
In this paper, instead, we focus mostly on the M\,101 ULS (often referred 
to in the literature as M\,101 ULX-1 or CXO\,J140332.3$+$542103), 
which has arguably the most extensive X-ray coverage among the ULS population. 

The dominant thermal component of ULS spectra has colour blackbody 
temperatures $kT_{\rm bb} \approx 50$--$150$ eV (hence, very little 
emission $> 1$ keV), often varying from observation to observation. 
The characteristic blackbody radii are 
$R_{\rm bb} \approx 10,000$--$100,000$ km. Blackbody-model bolometric 
luminosities reach $\approx$ a few $10^{39}$ erg s$^{-1}$, although 
for such low temperatures, luminosity estimates have to be taken 
with great caution. 
It is well known \citep{kahabka97,balman98,greiner00,ness08} that 
in Local Group supersoft sources, simple blackbody fits tend 
to overestimate the bolometric luminosity and underestimate the temperature, 
compared with white dwarf atmosphere models. 
It was also recently found \citep{ness13},
via X-ray grating spectroscopy, that the spectra of Local Group 
supersoft sources are far richer in absorption and emission lines 
than previously thought.
Unfortunately, for extragalactic ULSs, the count rate is too low 
for grating spectroscopy, and their physical properties have to be inferred 
from CCD-resolution spectroscopy.

Several models have been proposed for ULSs.
One suggestion \citep{kong03,kong04,liu08a} 
was that they are powered by intermediate-mass BHs 
(IMBHs) in a disk-dominated 
state. One problem of this scenario is that the observed 
colour temperatures are very low, often dipping below 100 eV. 
A standard disk reaches a peak temperature 
$kT_{\rm in} \approx 230 (\dot{m}/M_4)^{1/4}$ eV         
\citep{done12,soria07,kubota98}, 
where $M_4$ is the BH mass in units of $10^4 M_{\odot}$. 
The corresponding luminosity is $L \approx 1.3 \times 10^{42} \dot{m} M_4$ 
erg s$^{-1}$. Usually, an accreting BH is in a disk dominated state 
only for $0.02 \la \dot{m} \la 0.3$ \citep{maccarone03,steiner09} (canonical 
high/soft state), or more generally for $0.02 \la \dot{m} \la 1$ 
if we also include the non-standard disk regime near the Eddington limit. 
Combining this limit on $\dot{m}$ with the previous expressions 
for $kT_{\rm in}$ and $L$, it is easy to show that BHs are expected 
to be in a disk-dominated state only in a specific region 
of the temperature-luminosity plane, namely for 
\begin{equation}
5.2 \times 10^{38} \, \left(\frac{230{\mathrm{~eV}}}{kT_{\rm in}}\right)^4 
\la \frac{L}{{\mathrm{~erg~s}}^{-1}} 
\la 1.3 \times 10^{42} \left(\frac{230{\mathrm{~eV}}}{kT_{\rm in}}\right)^4.\label{lum_eq}
\end{equation}
As discussed in a companion paper \citep{urquhart15}, 
ULSs fall mostly outside of that region of parameter space. 
Therefore, a disk-dominated IMBH model is in most cases 
not self-consistent, although we cannot rule out the IMBH 
scenario a priori for all sources.


Alternatively, ULSs are extreme examples 
of quasi-steady surface-nuclear-burning white dwarfs in close binary systems, 
by analogy with (less luminous) supersoft sources in the Milky Way 
and Local Group 
\citep{vandenheuvel92,rappaport94,greiner00,greiner04,distefano04,orio10}.
The characteristic blackbody radii of ULSs are a few times larger 
than white dwarf radii. However, that is not inconsistent with the 
accreting white dwarf scenario: as the luminosity produced by shell burning 
reaches and exceeds the Eddington limit for a white dwarf 
($L \approx 10^{38}$ erg s$^{-1}$), we do expect envelope expansion 
and/or an optically thick outflow \citep{hachisu96,fabbiano03}.
Recurrent novae (also powered by nuclear burning 
on an accreting white dwarf) are known to exceed the Eddington limit 
in their outbursts: RS Oph reached $L \approx 10^{40}$ erg s$^{-1}$ 
and remained super-Eddington for at least two months during its 2006 outburst 
\citep{skopal15a,skopal15b}.

 
A third explanation for ULSs is that they are stellar-mass 
BHs or neutron stars accreting strongly above their Eddington limit, 
so that the central X-ray source is completely shrouded by 
a massive radiatively-driven outflow. The X-ray photons from the inner disk 
are downscattered and thermalized in the Compton-thick wind, 
and we are seeing the photosphere of the outflow 
\citep{mukai03,king03,fabbiano03,poutanen07,shen15}.
Mildly super-Eddington accretion is now the generally accepted explanation 
for the vast majority of standard ULXs. In this scenario, ULSs 
are the extreme end of the general ULX population, when viewed through 
the thickest winds (corresponding to a combination of highest accretion 
rates and sufficiently high viewing angle).

Recent multiband studies of the M\,101 ULS \citep{liu13} 
seemed to provide the solution for the nature of ULSs. However, 
such results have also posed some new unanswered questions: in particular, 
whether or not the soft emission is coming from a standard accretion disk, 
and if so, why it is so cold. 
In this paper, we discuss whether there is an alternative interpretation, 
that can better explain its X-ray spectral properties and high luminosity.

\begin{figure}
\begin{center}
\epsfig{figure=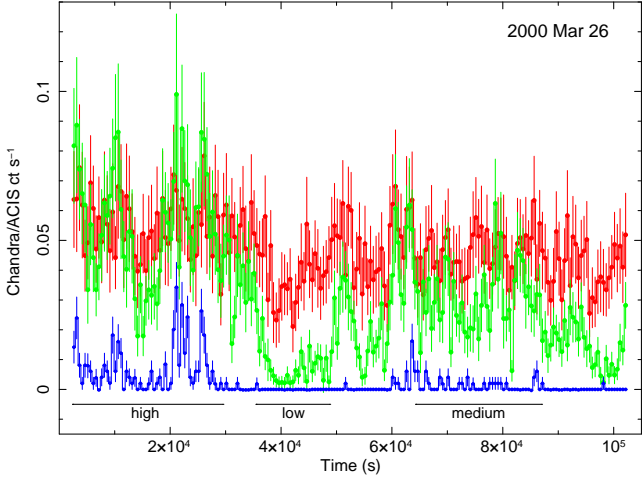,width=85mm,angle=0}
\end{center}
\caption{{\it {Chandra}}/ACIS-S lightcurve of the M\,101 ULS 
in the 0.3--0.7 keV (red datapoints), 0.7--1.5 keV (green datapoints) 
and 1.5--7 keV (blue datapoints) bands, in the 2000 March 26 observations, 
rebinned to 500s intervals. We also show the ``high'', ``medium'' 
and ``low'' sub-intervals previously defined and modelled by 
\citet{mukai03,kong05}. We use the same three sub-intervals 
in our spectral analysis, for a better comparison with the literature results.}
\label{fig1}
\end{figure}


\section{The ULS in M\,101: a challenge to standard disc models}

The {\it Chandra} position of the M\,101 ULS is 
R.A. $=$ 14$^{h}$03$^{m}$32$^{s}$.37, Dec. $=$ $+$54$^{\circ}$21'02''.8. 
Henceforth, we assume a Cepheid distance of 6.4 Mpc to M\,101 
(\citealt{shappee11}: $d = 6.4 \pm 0.2$ (random) $\pm 0.5$ (systematic) Mpc).
For comparison, a distance of 6.9 Mpc \citep{freedman01} 
was instead adopted by \cite{liu13}.
From their optical spectroscopic study, \cite{liu13} concluded  
that the M\,101 ULS contains a BH with mass $M_{\rm BH} > 5 M_{\odot}$, 
accreting from a Wolf-Rayet star of current mass $M_2 = (19 \pm 1) M_{\odot}$ 
(initial mass $M_{2,0} = (50 \pm 10) M_{\odot}$), in a binary system 
with orbital period $P = 8.24 \pm 0.1$ days.
With these parameters, the donor star is underfilling its Roche lobe: 
as a result, the system must be powered by wind accretion. In turn, this poses 
severe lower limits to the BH mass, in order for enough mass to be intercepted 
by the BH ($M_{\rm BH} \ga 50 M_{\odot}$ for a non-spinning BH). 
In such conditions, the M\,101 ULS cannot be a persistently 
super-Eddington source: 
in \cite{liu13}'s model, the accretion rate reaches only 
$\approx 10^{-7} M_{\odot}$ yr$^{-1}$: this is barely viable if the 
long-term-average bolometric luminosity of the BH 
is $\approx 3 \times 10^{38}$ erg s$^{-1}$, as opposed to a luminosity 
$\approx 3 \times 10^{39}$ erg s$^{-1}$ reached during bright states.

The supersoft thermal component is interpreted by \cite{liu13} as 
optically-thick disk emission, directly observed without any reprocessing 
in a hot corona. The main benefit of attributing the soft thermal emission  
directly to the disk is that this scenario can explain the luminosity variability 
between high- and low-flux epochs. The stellar wind itself cannot change 
by two orders of magnitude over few weeks; however, if accretion is mediated 
by a disk, thermal/viscous instabilities may give rise to the observed 
spectral transitions.  
This introduces two additional constraints on the BH mass. Firstly, 
the disk cannot form in a wind-accreting system unless 
the circularization radius of the captured wind is larger than 
the innermost stable circular orbit: for this to occur, the BH mass 
must be $M_{\rm BH} \ga 80 M_{\odot}$ or $M_{\rm BH} \ga 50 M_{\odot}$, depending 
on which stellar wind model is adopted \citep{liu13}. Secondly, a standard accretion disk 
is in the thermal dominant state only below $\approx 30$\% of 
the Eddington luminosity, {\it i.e.}, $\dot{m} \la 0.3$   
\citep{fender04,mcclintock06,steiner09}. Coupled with the inferred bolometric 
luminosity in the bright states, this constraint on the Eddington 
ratio requires the BH mass to be $M_{\rm BH} \ga 80 M_{\odot}$, possibly 
inconsistent with a stellar origin of the BH. 
Finally, no stellar-mass BH has ever 
been observed with a pure disk spectrum at temperatures $< 0.1$ keV: 
as \cite{liu13} note, their own interpretation of this ULS  
challenges standard models of BH accretion and disk structure.
Considering the unsolved problems of the disk scenario, 
it is worth re-examining alternative explanations for the supersoft 
thermal component, which we will discuss in this paper.

\begin{figure*}
\begin{center}
\epsfig{figure=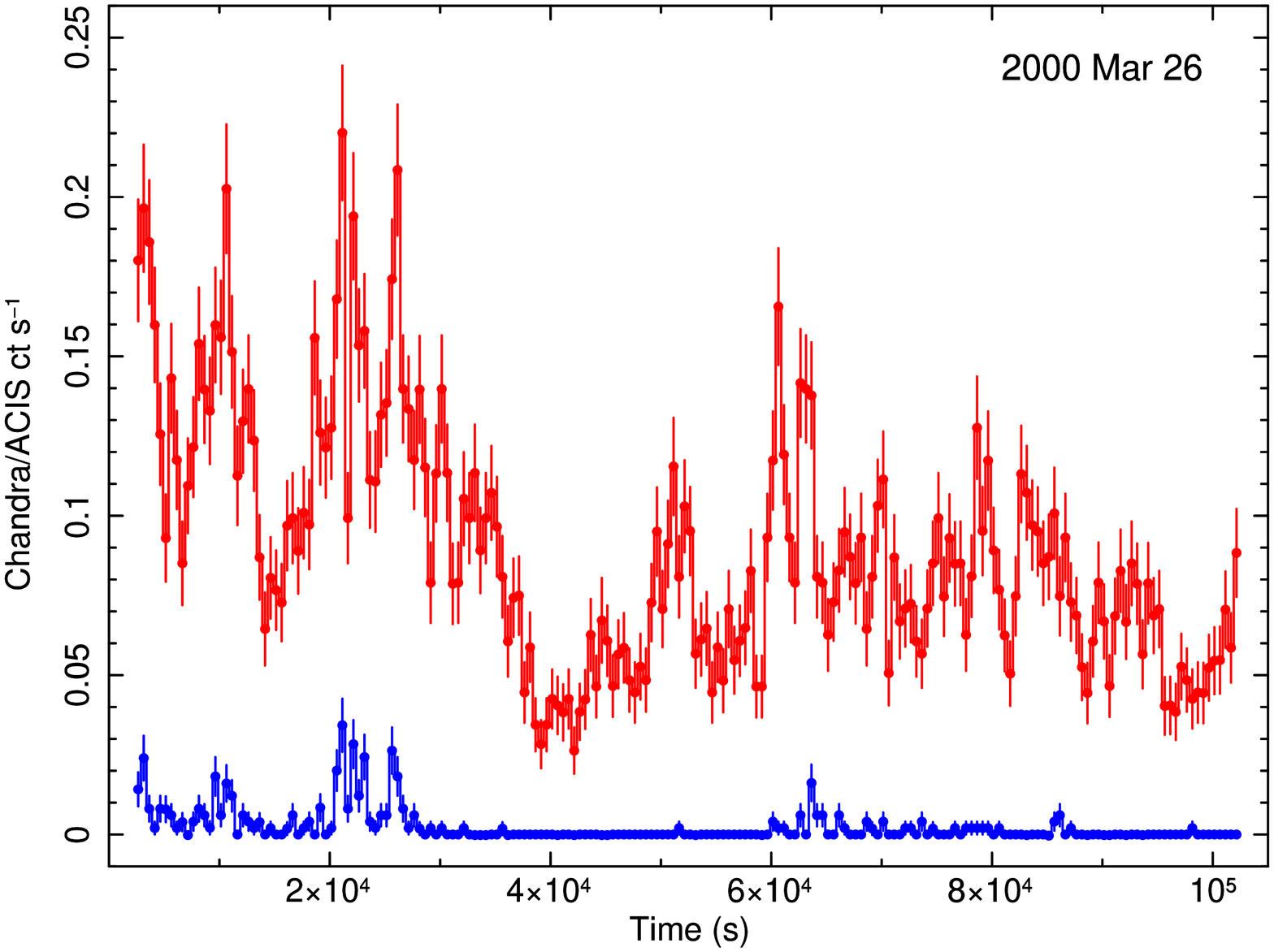,width=85mm,angle=0}
\hspace{0.2cm}
\epsfig{figure=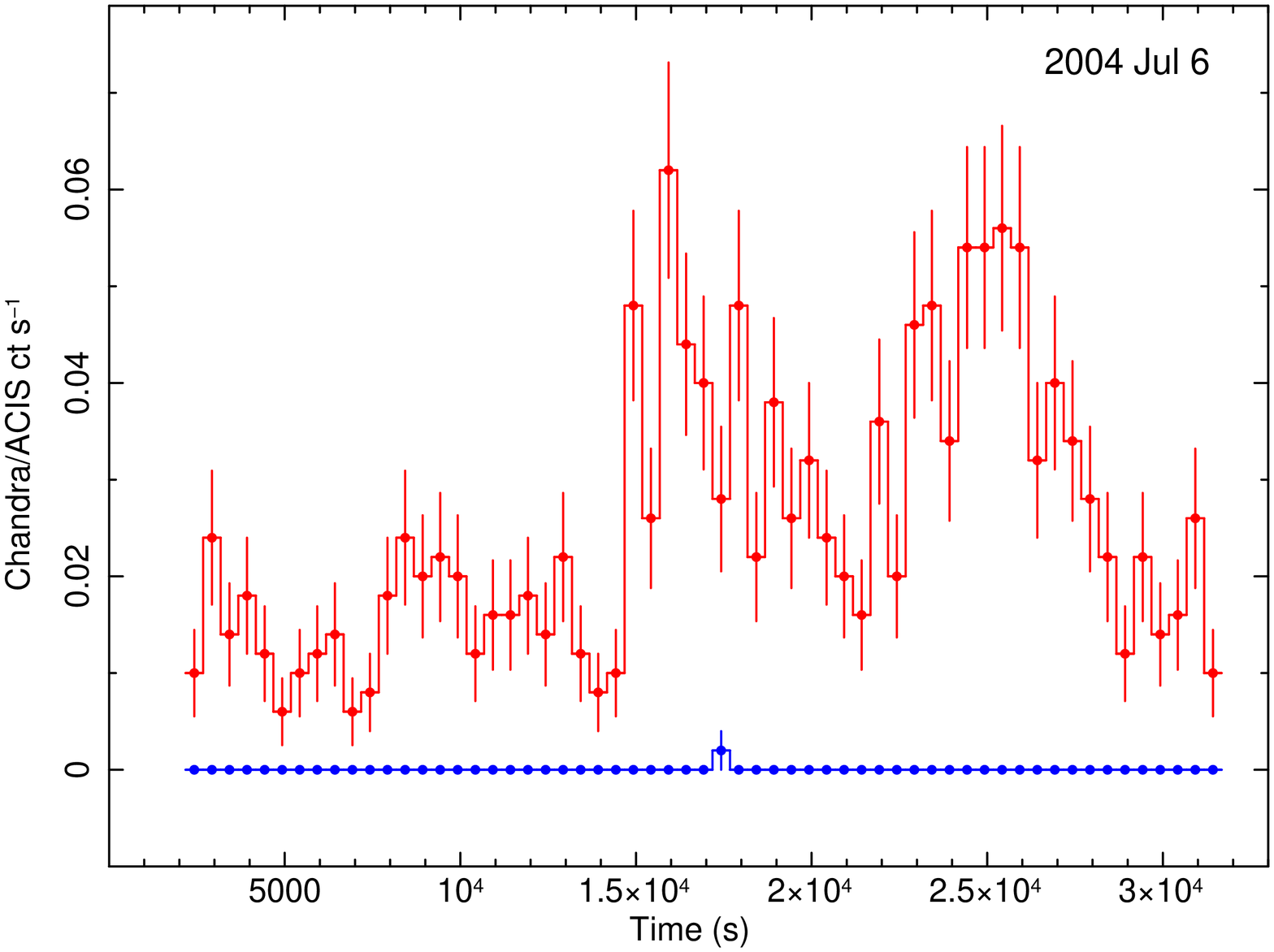,width=85mm,angle=0}\\[5pt]
\epsfig{figure=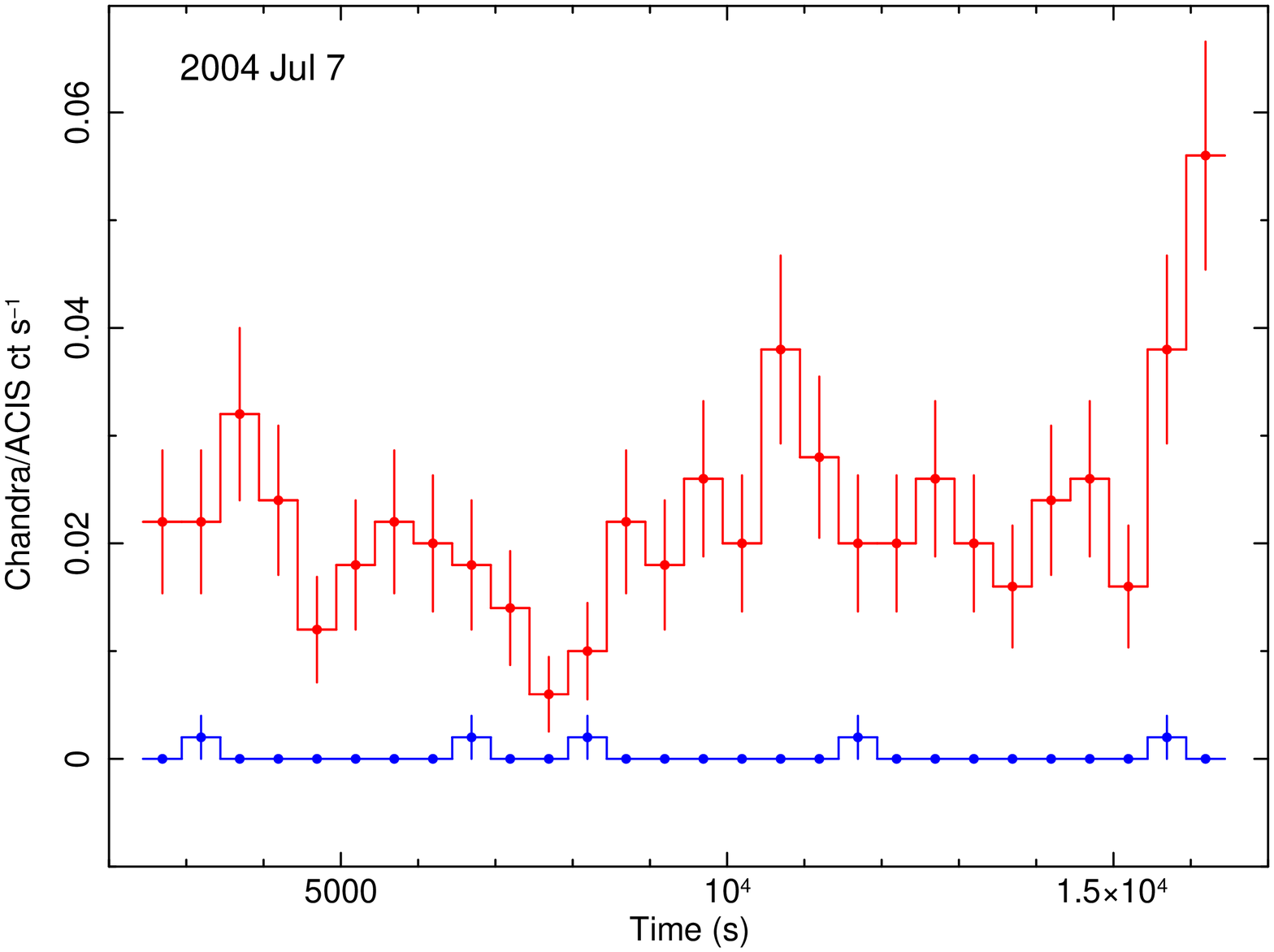,width=85mm,angle=0}
\hspace{0.2cm}
\epsfig{figure=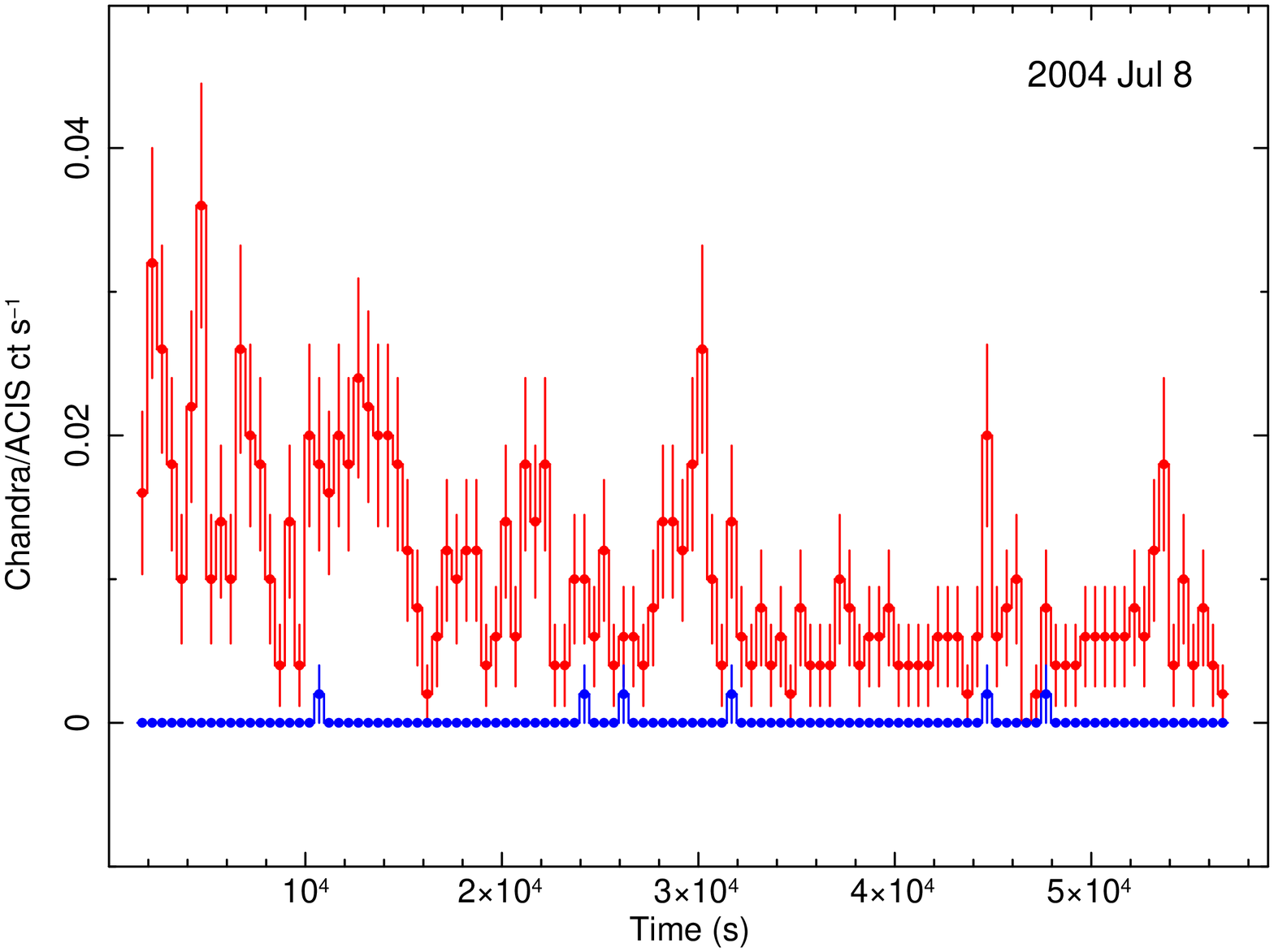,width=85mm,angle=0}\\[5pt]
\psfig{figure=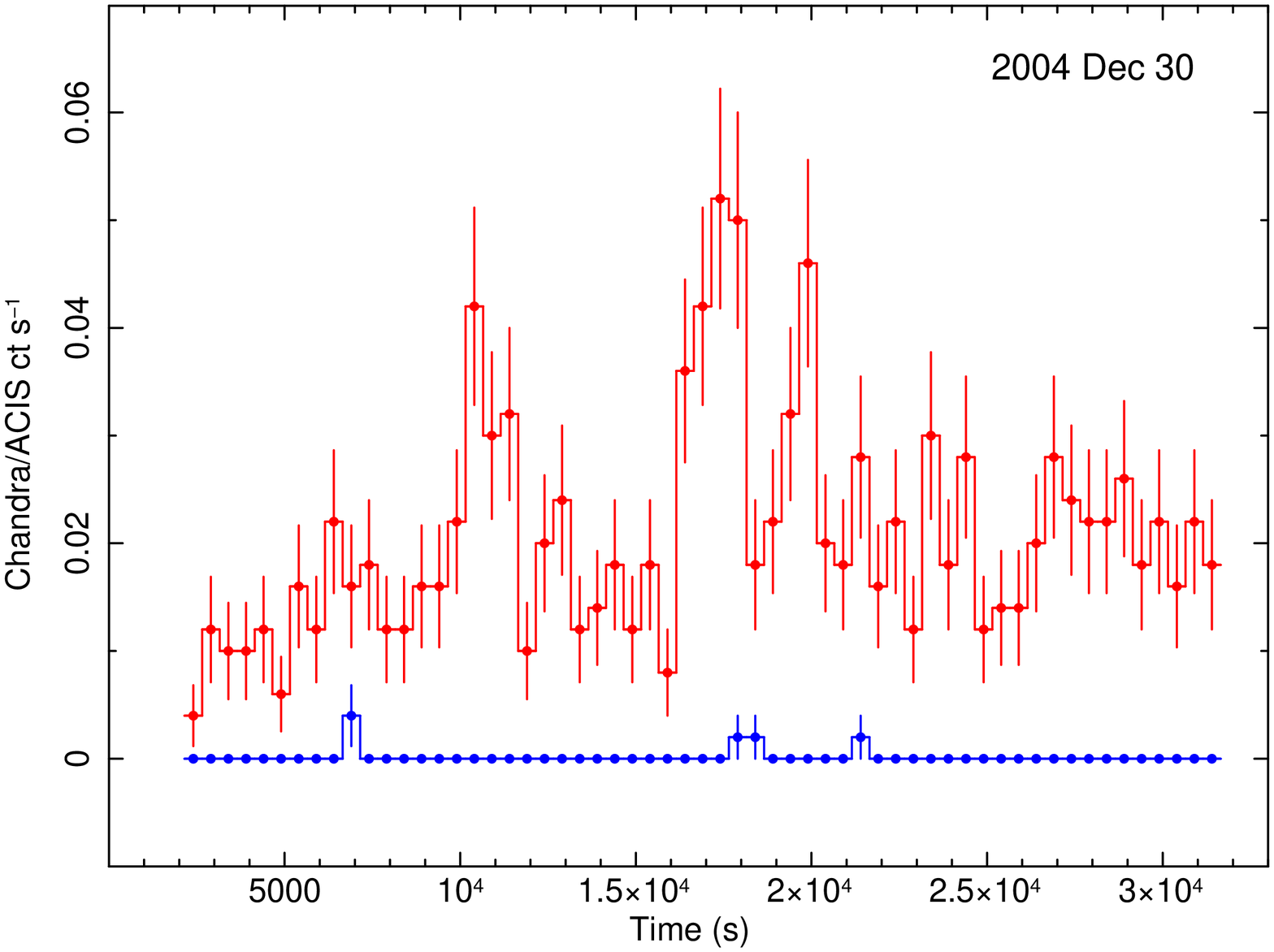,width=85mm,angle=0}
\hspace{0.2cm}
\psfig{figure=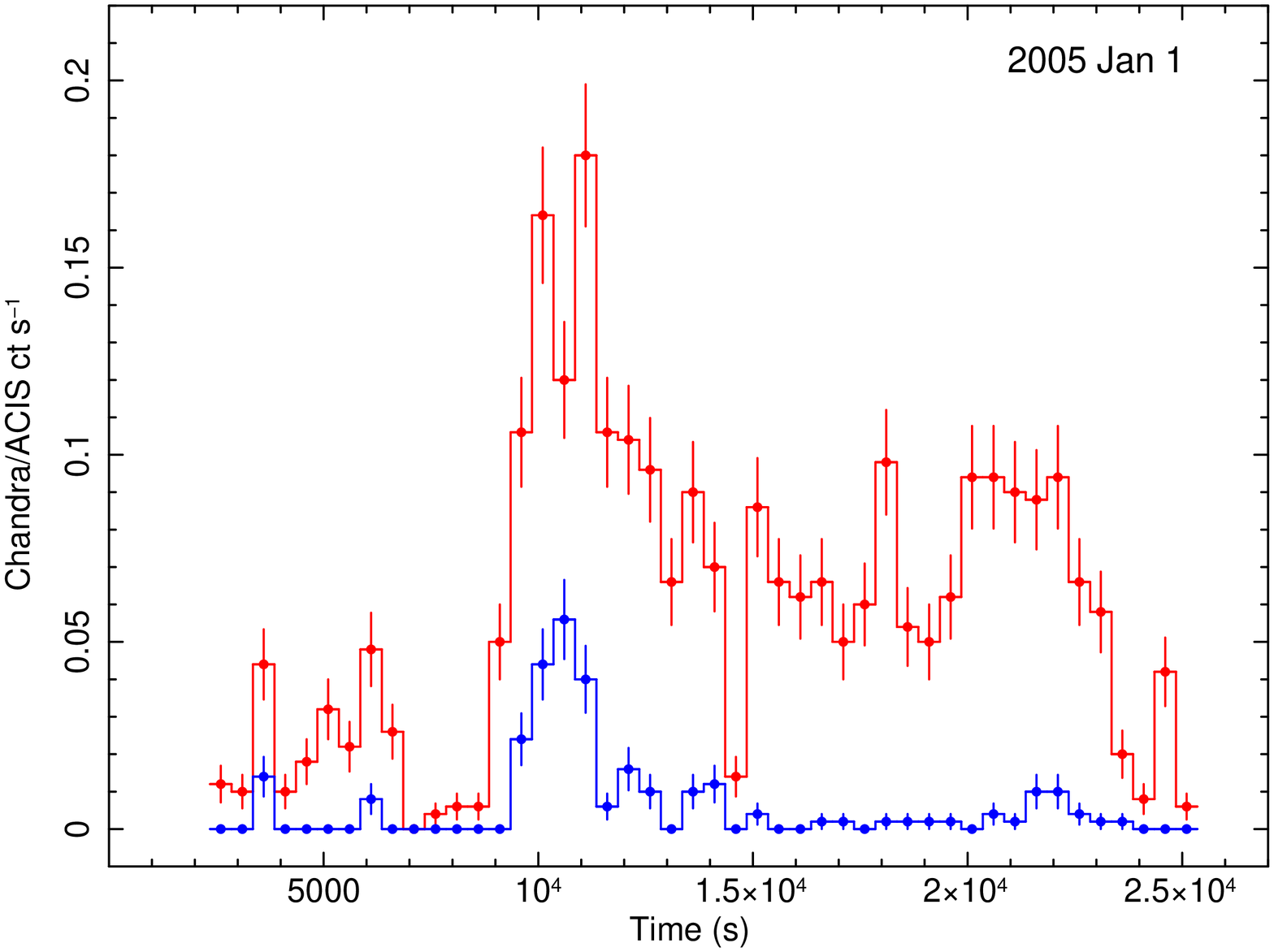,width=85mm,angle=0}
\end{center}
\caption{{\it {Chandra}}/ACIS-S lightcurve of the M\,101 ULS 
in the 0.3--1.5 keV band (red datapoints) and 1.5--7 keV band
(blue datapoints), for selected observations. All lightcurves 
are rebinned to 500s intervals.}
\label{fig2}
\end{figure*}

\section{Data analysis}
M\,101 has been observed by many X-ray missions over the years. 
For this paper, we have chosen to re-analyze the 25 {\it Chandra} observations, from 2000 March--October and 2004 January--2005 January, and 
the two {\it XMM-Newton} observations from 2004 July and 2005 January. 
All of them are available on their respective public archives.
There is also another {\it XMM-Newton} observation from 2002 June in which 
the source is barely detected at 3-$\sigma$ level \citep{kong04,jenkins04} 
but with insufficient signal-to-noise ratio to provide any useful 
spectral or colour information; therefore, we did not use the 2002 June 
observation in this paper. 
The ULS was also previously detected in three of the 12 
{\it ROSAT}/High Resolution Imager observations taken between 1992 and 1996 
(source H32 in \citealt{wang99}); however, the combined number of counts 
from those three (short) observations is only $\approx 50$. Therefore, 
a full re-analysis of the {\it ROSAT} data does not provide particularly 
useful additional information.


\begin{table*}
\caption{Root-mean-square fractional variability in different energy bands, for selected observations with high signal-to-noise ratio.}
\begin{center}
\begin{tabular}{lrrrr}
\hline\hline\\[-7pt]
Epoch & Frequency band (Hz) &  rms (0.3--0.7 keV) & rms (0.7--1.5 keV) & rms (1.5--7.0 keV) \\[-1pt]
\hline\\[-9pt]
2000 Mar 26 & $10^{-5}$--0.05 & $< 36\%$ & $(70\pm4)\%$ & $(100\pm24)\%$ \\[3pt]
2000 Oct 29 & $1 \times 10^{-4}$--0.01 & $(31\pm15)\%$ & -- & -- \\[3pt]
2004 Jul 06 & $4 \times 10^{-5}$--0.05 & $(41\pm17)\%$ & $(76\pm29)\%$ & -- \\[3pt]
2004 Jul 08 & $2 \times 10^{-5}$--0.05 & $(102\pm11)\%$ & -- & -- \\[3pt]
2004 Jul 11 & $3 \times 10^{-5}$--0.01 & $(50\pm19)\%$ & -- & -- \\[3pt]
2004 Dec 30 & $3 \times 10^{-5}$--0.01 & $(22\pm17)\%$ & $(63\pm21)\%$ & -- \\[3pt]
2005 Jan 01 & $5 \times 10^{-5}$--0.05 & $(58\pm12)\%$ & $(89\pm6)\%$ & $(162\pm17)\%$\\[3pt]
2005 Jan 08 & $7 \times 10^{-5}$--0.01 & $(27\pm10)\%$ & -- & -- \\
\hline\\[-12pt]
\end{tabular}
\end{center}
\label{tab2}
\end{table*}

For {\it Chandra}, we re-processed the data  
with standard tasks in the {\small CIAO} Version 4.6 \citep{fruscione06} 
data analysis system. In particular, we used the {\small CIAO} task 
{\it specextract} to extract a spectrum (with its associated 
background, response and ancillary response files) from each 
observation. In a few cases (Table 1), there are enough counts for 
a meaningful spectral modelling, and the observed count rate 
corresponds to X-ray luminosities $\sim$$10^{39}$ erg s$^{-1}$. 
In most other cases, the ULS is barely detected at $\sim$$3\sigma$ 
level, corresponding to X-ray luminosities $\sim$$10^{37}$ erg s$^{-1}$. 
We carried out individual spectral fitting for the data from 
ObsIDs 934, 2065, 4737, 6169. In fact, we split the long ObsID 934 (98 ks) 
into three intervals with high, medium and low observed count rates, 
defined exactly as in \cite{mukai03} and \cite{kong05}, and fitted them individually. (The three sub-intervals are illustrated in Figure 1, and discussed 
in more detail in Section 4.1.)
We combined and fitted the average 
spectrum from ObsIDs 4734, 5337, 5338, 5339 and 5340 (spanning 
the time range 2004 July 5--11, for a total exposure time of 143 ks), 
and the average spectrum 
from ObsIDs 6170 and 6175 (2004 December 22--24; 89 ks).
Finally, we produced and fitted a deep combined spectrum 
from the 14 ``faint-state'' observations in which the source 
had a 0.3--2 keV count rate $< 5 \times 10^{-4}$ ct s$^{-1}$ 
(ObsIDs 4731, 5297, 5300, 5309, 4732, 5322, 4733, 5323, 6114, 6115, 
6118, 4735, 4736 and 6152, for a total exposure time of 670 ks).
All combined spectra were produced with {\it specextract}, 
so that individual spectral and response files were created 
for each epoch and then averaged.

For both the {\it XMM-Newton} observations considered here, we reprocessed the Observation Data Files with the Science Analysis System (SAS) version 14.0.0 (xmmsas\_20141104). For the 2004 July 23 observation, we had to remove 
about 1/3 of the exposure time due to background flaring.
We extracted the source events from a circular region centred on the ULS, with a 20'' radius; we extracted the local background from a region three times as large, suitably selected to avoid any other bright sources or chip gaps. 
We selected single and double events (pattern $\le 4$ for the pn and pattern $\le 12$ for MOS1 and MOS2), with the standard flagging criteria \#XMMEA\_EP and \#XMMEA\_EM for the pn and MOS, respectively, and with the stricter flagging condition FLAG=0 for spectral analysis. We built response and ancillary response files with the SAS tasks {\it rmfgen} and {\it arfgen}, and we then created an average EPIC spectrum and response file with {\it epicspeccombine}. We grouped the combined spectrum to a minimum of 20 counts per bin, for $\chi^2$ fitting. For the 2005 January 8 observation, we also fitted the pn and MOS spectra simultaneously, verifying that we obtained a result consistent with that obtained from the combined spectrum within the 90\% confidence limit; 
for the 2004 July 23 observation, there are not enough counts in the MOS to permit simultaneous spectral fitting.  

We used {\small XSPEC} Version 12.6 \citep{arnaud96} for spectral fitting, 
both for {\it Chandra} and {\it XMM-Newton}.   
For models involving pileup (see Section 4.3), 
we also double-checked the results 
with the spectral fitting package {\small ISIS} \citep{houck00},
and found them consistent within the 90\% uncertainties. 
Both {\small XSPEC} and {\small ISIS} use independent implementations 
of a pile-up model developed by \cite{davis01}.

For the time variability study, we used the {\small CIAO} task 
{\it dmextract} to extract {\it Chandra} lightcurves 
in the soft (0.3--1.5 keV) and hard (1.5--7 keV) band. 
For the {\it XMM-Newton} observations, we used the {\small SAS} tasks 
{\it xmmselect} and {\it epiclccorr} to extract background-subtracted 
EPIC-pn and EPIC-MOS lightcurves. 
We then used standard {\small FTOOLS} tasks for timing analysis and statistics 
\citep{blackburn95}. 



\section{Results}

\subsection{Time variability}  

The most striking feature of the X-ray lightcurves at most epochs 
is their strong short-term variability (Figures 1,2), as already noted 
in the literature \citep{mukai03,kong05}. 
The observed flux often varies by a factor of 3 in $\approx$$10^3$ s. 
This is much faster than any $e$-folding rise or decay timescale 
of transient Galactic X-ray binaries attributed to disk instabilities 
(typically, $\ga 1$ d). The root-mean-square (rms) 
fractional variability \citep{edelson02,markowitz03,vaughan03,gierlinski05,middleton11} of the 0.3--1.5 keV lightcurve, for frequencies 
$\le 0.1$ Hz, is as high as $(32 \pm 3)\%$ on 2000 March 26, 
$(58 \pm 10)\%$ on 2004 July 6, $(100 \pm 10)\%$ on 2004 July 8,
and $(71 \pm 5)\%$ on 2005 January 1.
For the observations with the highest signal-to-noise ratio, 
we also computed the rms fractional variability separately for 
the 0.3--0.7 keV and 0.7--1.5 keV bands (Table 2).
The behaviour of the source on 2000 March 26 and 2005 January 1 
suggests that the rms variability increases at harder energies (Table 2).

Modelling the physical origin of this variability is beyond the scope 
of this paper. Here, we just compare the level of rms variability 
seen in the M\,101 ULS with that seen in other accreting BHs. 
The observed rms variability $\sim$30\%--100\% is higher than 
in any canonical state of stellar-mass BHs \citep{belloni10}, 
and certainly inconsistent with the disk-dominated 
high/soft state, when the rms is $< 10\%$. This is another 
strong argument against an IMBH disk model for this ULS.


Instead, high rms variability (up to $\sim$50\%) and an increase 
in rms variability 
at higher energy bands are two characteristic properties of ULXs 
in the soft-ultraluminous regime \citep{middleton11,sutton13,middleton15}. 
This has been interpreted as due to variable 
obscuration of the hard inner-disk emission by a clumpy disk wind, 
for ULXs seen at high inclination angle.
Based on this analogy, we speculate that the M\,101 source and other ULSs 
might be an extreme case of the soft-ultraluminous regime. In this scenario, 
ULSs could be super-critical accreting sources in which a clumpy outflow 
almost completely masks and reprocesses the harder emission from 
the central regions. The outflow must have an even higher optical depth 
than in the soft-ultraluminous sources discussed by \citet{sutton13}.
We will outline a possible analytic model for such an outflow in Section 5.2.  


In most epochs, there is either no significant detection in the hard band 
(1.5--7 keV), or it is consistent with a count rate $\la 10^{-4}$ ct s$^{-1}$.
Significant spikes in the hard flux are seen (Figures 1,2) during the strongest 
soft flares of ObsID 934 (2000 March 26) and ObsID 4737 (2005 January 1). 
On both occasion, the hard emission appears only when the X-ray source 
reaches count rates $\ga 0.1$ ct s$^{-1}$.
We will use spectral analysis (Sections 4.2 and 4.3) to determine whether 
the count-rate variability is associated to real changes in the emission 
properties or (for example) to occultation. We will also discuss whether 
the hard flares 
are an additional physical component, or are the Wien tail of the supersoft 
thermal component, or are entirely due to photon pileup, which is known 
to be significant at count rates $\ga 0.1$ ct s$^{-1}$.


\begin{figure*}
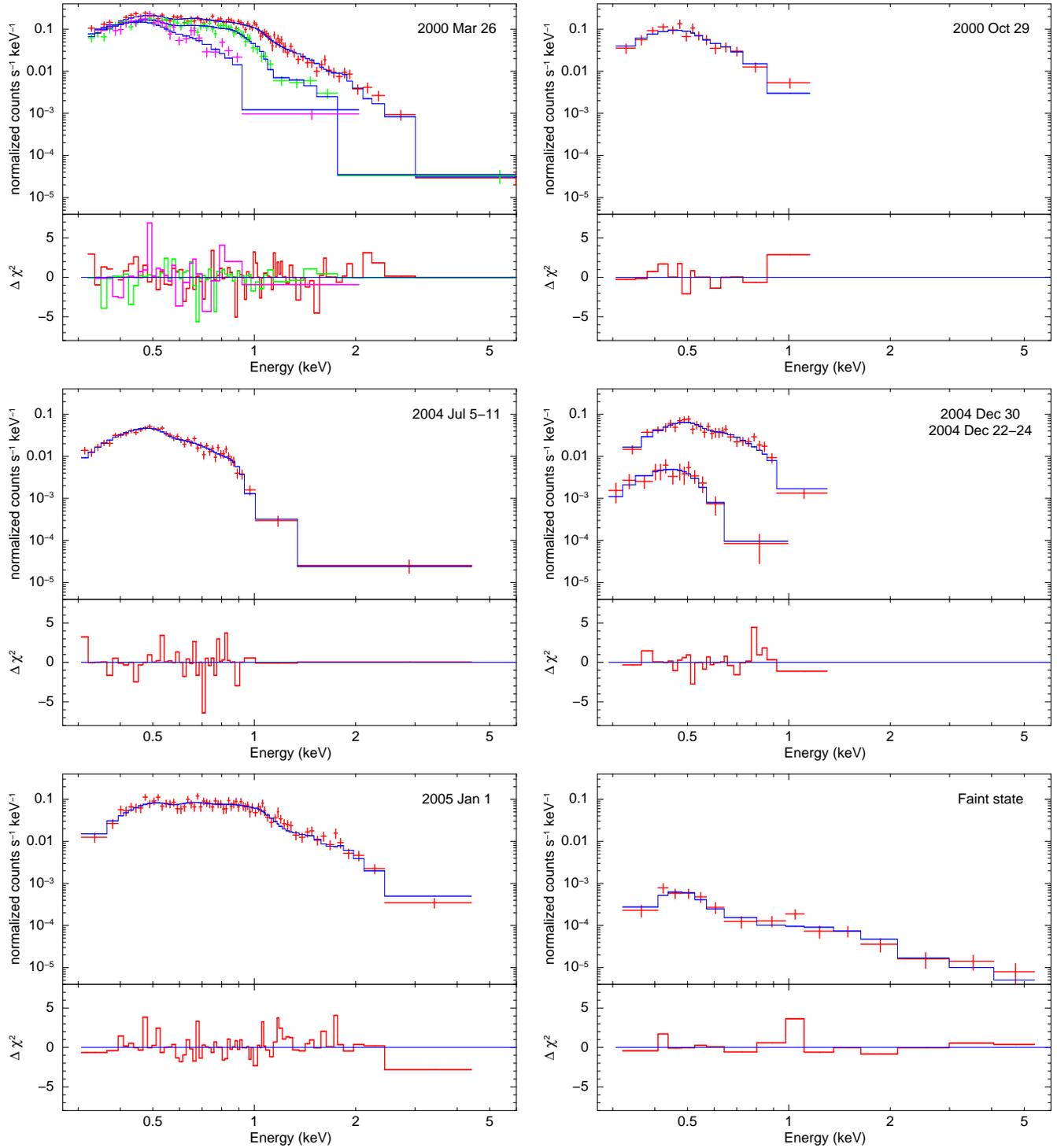

\begin{center}
\epsfig{figure=fig3a.ps,width=63mm,angle=270}
\hspace{0.2cm}
\epsfig{figure=fig3b.ps,width=63mm,angle=270}\\[5pt]
\epsfig{figure=fig3c.ps,width=63mm,angle=270}
\hspace{0.2cm}
\epsfig{figure=fig3d.ps,width=63mm,angle=270}\\[5pt]
\psfig{figure=fig3e.ps,width=63mm,angle=270}
\hspace{0.2cm}
\psfig{figure=fig3f.ps,width=63mm,angle=270}
\end{center}
\caption{{\it {Chandra}}/ACIS-S spectral data and $\chi^2$ residuals 
at different epochs. The full set of model components and parameters used 
for these plots are listed 
in Tables A2 (for ObsID 934 = 2000 March 26), and Table A4 (for all other 
epochs). The best-fitting parameters of the dominant blackbody component 
at each epoch are also summarized in Table 3.   
Datapoints have been binned to $> 15$ counts per bin. 
For ObsID 934 (top left panel), we split the spectrum 
into a high-count-rate (red datapoints and residuals), medium-count-rate 
(green) and low-count-rate (magenta) sub-intervals.
We did not plot the residuals from the stacked observations of 
2004 December 22--24 
because the corresponding fit was obtained with the Cash statistics rather 
than with $\chi^2$; the stacked spectrum of 2004 December 22--24 
was then rebinned 
to a signal-to-noise ratio $\ge 2$ for display purposes only.}
\label{fig3}
\end{figure*}

\begin{figure*}
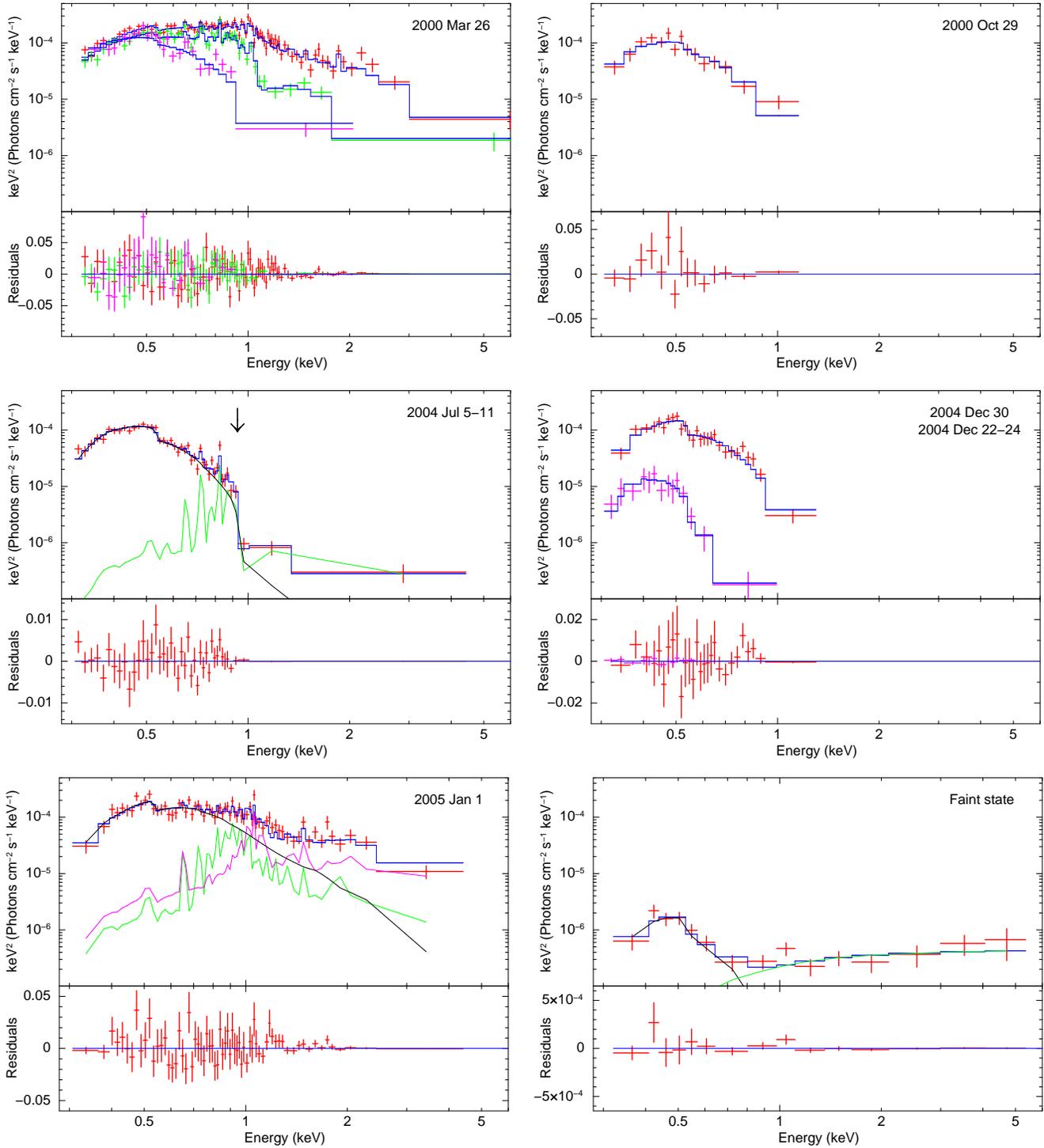

\begin{center}
\epsfig{figure=fig4a.ps,width=63mm,angle=270}
\hspace{0.2cm}
\epsfig{figure=fig4b.ps,width=63mm,angle=270}\\[5pt]
\epsfig{figure=fig4c.ps,width=63mm,angle=270}
\hspace{0.2cm}
\epsfig{figure=fig4d.ps,width=63mm,angle=270}\\[5pt]
\psfig{figure=fig4e.ps,width=63mm,angle=270}
\hspace{0.2cm}
\psfig{figure=fig4f.ps,width=63mm,angle=270}
\end{center}
\caption{Same set of {\it {Chandra}}/ACIS-S spectra as shown in Figure 3, 
but this time plotted as unfolded spectra ({\it eeuf} in {\small XSPEC}, 
corresponding to units of $\nu f_{\nu}$) and count-rate residuals. 
When the best-fitting model includes multiple components, they are also 
plotted in each panel, except for the 2000 March 26 spectrum 
(top left panel) which is 
expanded for clarity in Figure 5. An absorption edge in the stacked 
2004 July 5--11 spectrum (middle row, left panel) is marked with a black arrow.
The full set of model parameters corresponding to these plots are listed 
in Tables A2, A4. Datapoints have been binned to $> 15$ counts per bin.}
\label{fig4}
\end{figure*}

\begin{figure}
\begin{center}
\epsfig{figure=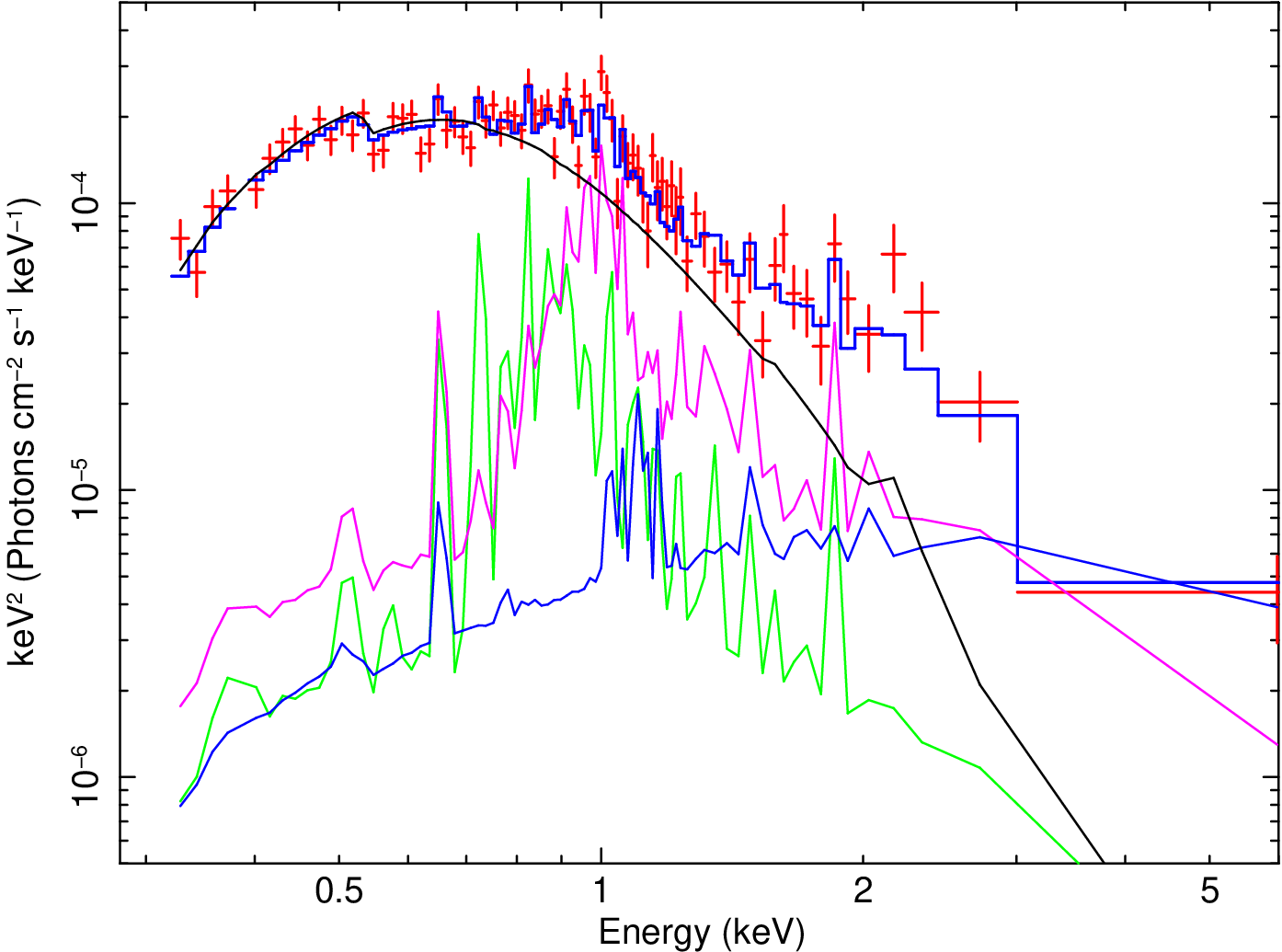,width=62mm,angle=270}\\[5pt]
\epsfig{figure=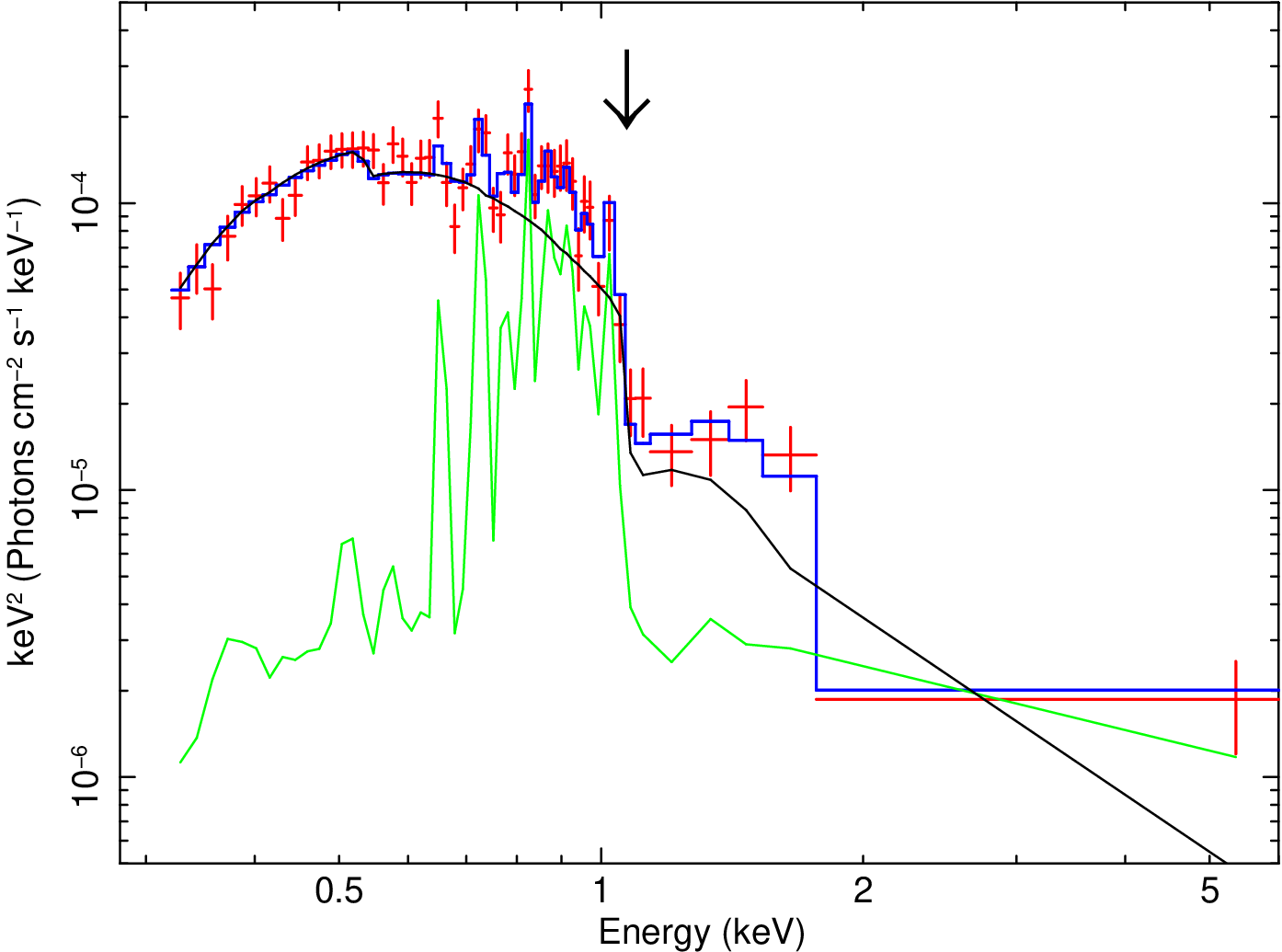,width=62mm,angle=270}\\[5pt]
\epsfig{figure=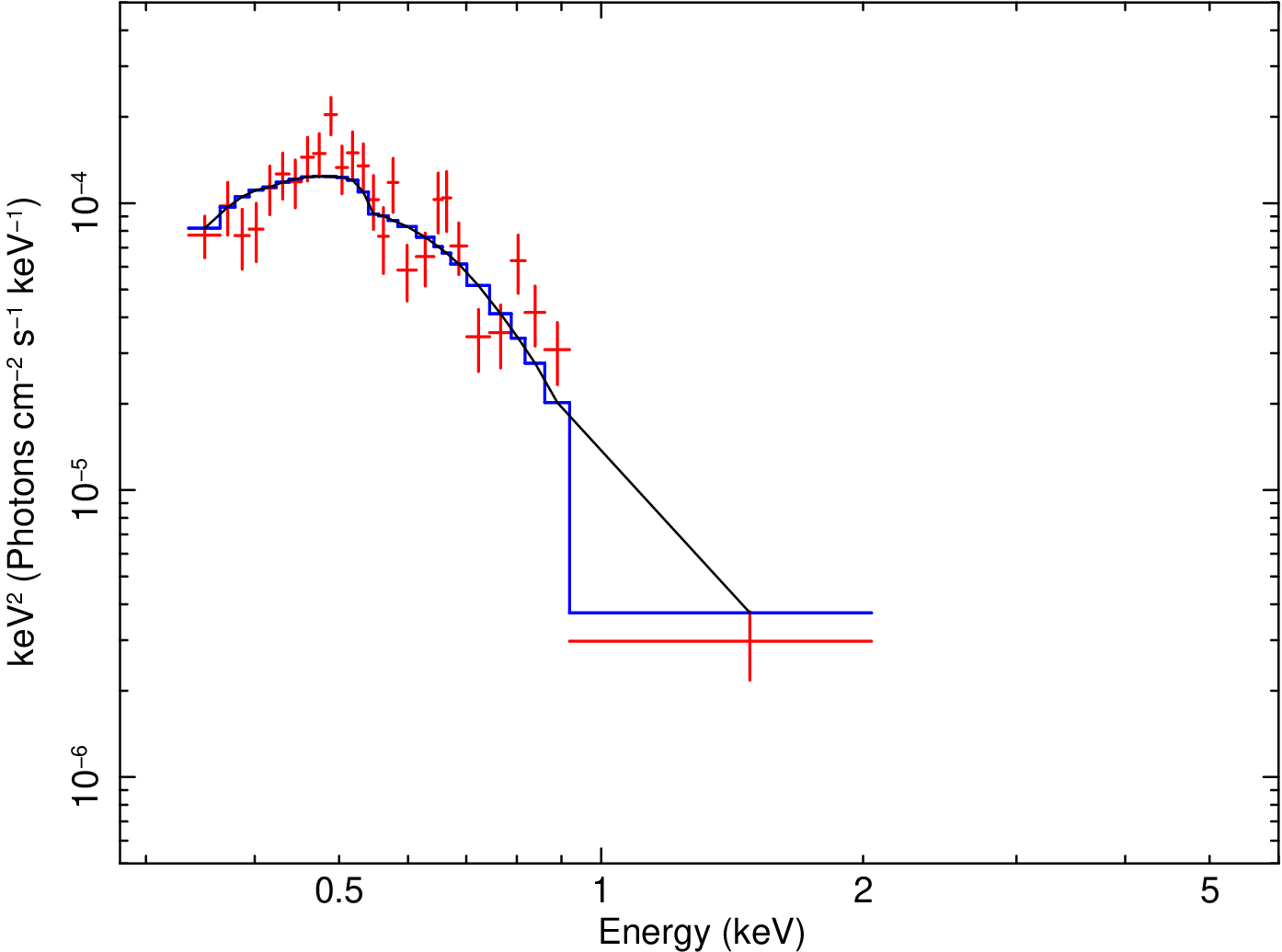,width=62mm,angle=270}
\end{center}
\caption{Unfolded spectra and model components 
for the {\it {Chandra}} ObsID 934, 
split into high, medium- and low-count-rate sub-intervals 
(as defined in Figure 1; 
see also the top left panel of Figures 3,4). The absorbing column density 
and the temperatures (but not the normalizations) of the three {\it mekal} 
components were kept locked for the three sub-intervals. See Table A2 
for the best-fitting parameters.
Top panel: the spectrum from the high-count-rate interval 
has three significant {\it mekal} components in addition to 
the soft blackbody. Middle panel: the medium-count-rate spectrum 
has only one significant {\it mekal} component, but it has 
an absorption edge (marked by the black arrow). 
Bottom panel: the low-count-rate spectrum 
is well fitted by a simple absorbed blackbody, with no hard excess 
or other residuals.}
\label{fig5}
\end{figure}

\subsection{Optically-thick thermal continuum}

All spectra are dominated (Figures 3,4 and Appendix A) 
by a soft thermal component with 
a characteristic temperature $\la 0.1$ keV, therefore with 
a peak flux at photon energies $\la 0.3$ keV, at 
the lowest energy range of the {\it Chandra}/ACIS-S 
and {\it XMM-Newton}/EPIC detectors. 
In addition, some spectra have additional 
harder emission detected at $\sim$1--5 keV, contributing 
a few per cent of the total luminosity. 
The combined spectrum of the faint-state observations 
(bottom right panel in Figures 3,4)
is dominated by a power-law-like component but it also has 
a significant supersoft peak, consistent with a temperature 
of $\approx$50 eV, thus emitting almost entirely below 
the {\it Chandra} band. The best-fitting simple power-law model 
for the stacked faint-state spectrum has $\chi^2 = 39.8$ for 12 degrees 
of freedom, while adding an $\approx$50-eV blackbody component 
significantly improves the fit down to $\chi^2 = 9.8$ for 10 degrees 
of freedom. 
In this section, we discuss our interpretation of the supersoft 
thermal continuum, which is the main focus of this work. 
In Section 4.3, we will discuss evidence and interpretations  
for the harder components. 


Because of the very low temperature of the thermal emission, 
we find (unsurprisingly) that it is impossible 
to distinguish between a disk-blackbody and a single-temperature 
blackbody model. We are seeing only the Wien part of the thermal 
spectrum, which is almost identical in the two models. 
The only difference is that for each spectrum, the fitted colour 
temperature in the disk-blackbody model is slightly higher than the 
corresponding simple blackbody temperature (cf.~Tables 3 and 4, 
and Model 4 versus Model 5 in Table A1), because 
of the slightly different way temperatures and normalization 
are defined in the two models. We verified that blackbody and disk-blackbody 
models are statistically equivalent for every epoch. For the rest of the paper, 
unless specifically indicated, we will use radii and temperatures 
from the blackbody fits, to allow a direct comparison with the outflow model. 

Our fits show that the bolometric luminosity 
is always $\ga 2 \times 10^{39}$ erg s$^{-1}$ except 
for the faint state (Tables 3, 4). In the faint state, the extremely low 
temperature of the soft component makes it impossible to estimate 
a reliable bolometric luminosity (most of the emission being 
in the far-UV), but we cannot rule out (Table 3) that it is 
also consistent with $\ga 2 \times 10^{39}$ erg s$^{-1}$.
In fact, despite the huge differences (three orders of magnitude) 
in the observed X-ray count rates, all observations are consistent 
with a blackbody bolometric luminosity 
varying only within a small range of values, 
between $\approx$ 3--10 $\times 10^{39}$ erg s$^{-1}$.
We have already mentioned (Section 1) that such bolometric 
extrapolations in supersoft sources may overestimate 
the true luminosity. Even if that is the case, 
the general significance of our result is that 
the large difference in X-ray count rates
between ``bright'' and ``faint'' states 
is mostly due to changes in the blackbody temperature, 
causing the thermal component to slip out of 
the detector sensitivity. We suggest that this is analogous 
to the well-documented behaviour of some classical supersoft sources, 
which can switch between a UV-bright phase (when the optically-thick 
envelope expands and cools) and an X-ray-bright phase 
(when the radius of the envelope decreases and its photospheric 
temperature increases) \citep{vandenheuvel92,pakull93}.  
We find no significant trend of the bolometric luminosity versus 
fitted temperature or radius, which is to say that $R_{\rm bb}$
roughly scales as $T_{\rm bb}^{-2}$ (as shown in more details in Section 5.3). 


\begin{figure}
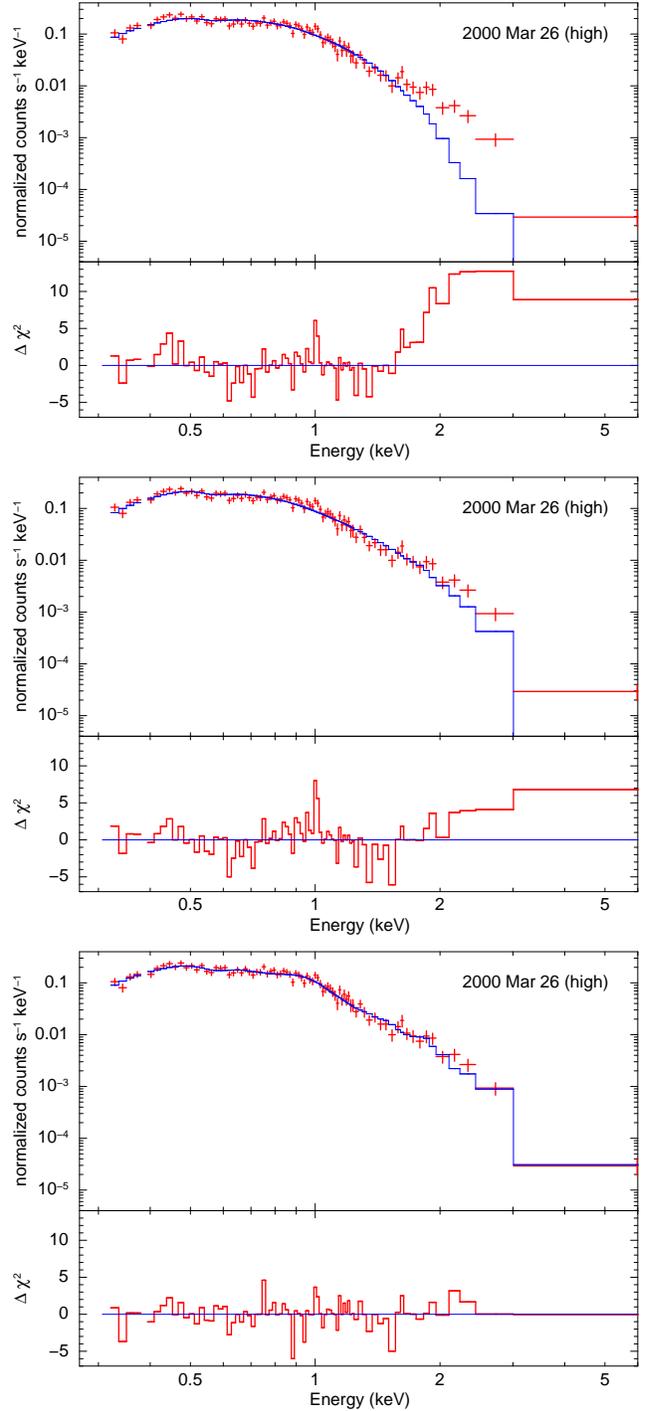

\begin{center}
\epsfig{figure=fig6a.ps,width=61mm,angle=270}\\[5pt]
\epsfig{figure=fig6b.ps,width=61mm,angle=270}\\[5pt]
\epsfig{figure=fig6c.ps,width=61mm,angle=270}
\end{center}
\caption{{\it {Chandra}}/ACIS-S spectral data and $\chi^2$ residuals 
for the high-count-rate interval of ObsID 934, fitted with three different 
models.
Top panel: absorbed single-temperature blackbody, showing a significant 
excess $> 1.5$ keV (Model 1 in Table A1; $\chi^2_{\nu} = 176.2/79$).  
Middle panel: absorbed single-temperature blackbody convolved with 
a pileup model; although 
some of the hard photons are now accounted for, the shape of the model 
is still unsatisfactory (Model 2 in Table A1; $\chi^2_{\nu} = 128.7/78$). 
Bottom panel: absorbed piled-up blackbody plus two-temperature 
thermal plasma emission ({\it mekal} model), providing a significantly 
better fit with $\chi^2_{\nu} = 84.9/74$ (Model 4 in Table A1). 
Datapoints have been binned to $> 15$ counts per bin before fitting.}
\label{fig6}
\end{figure}

\begin{figure}
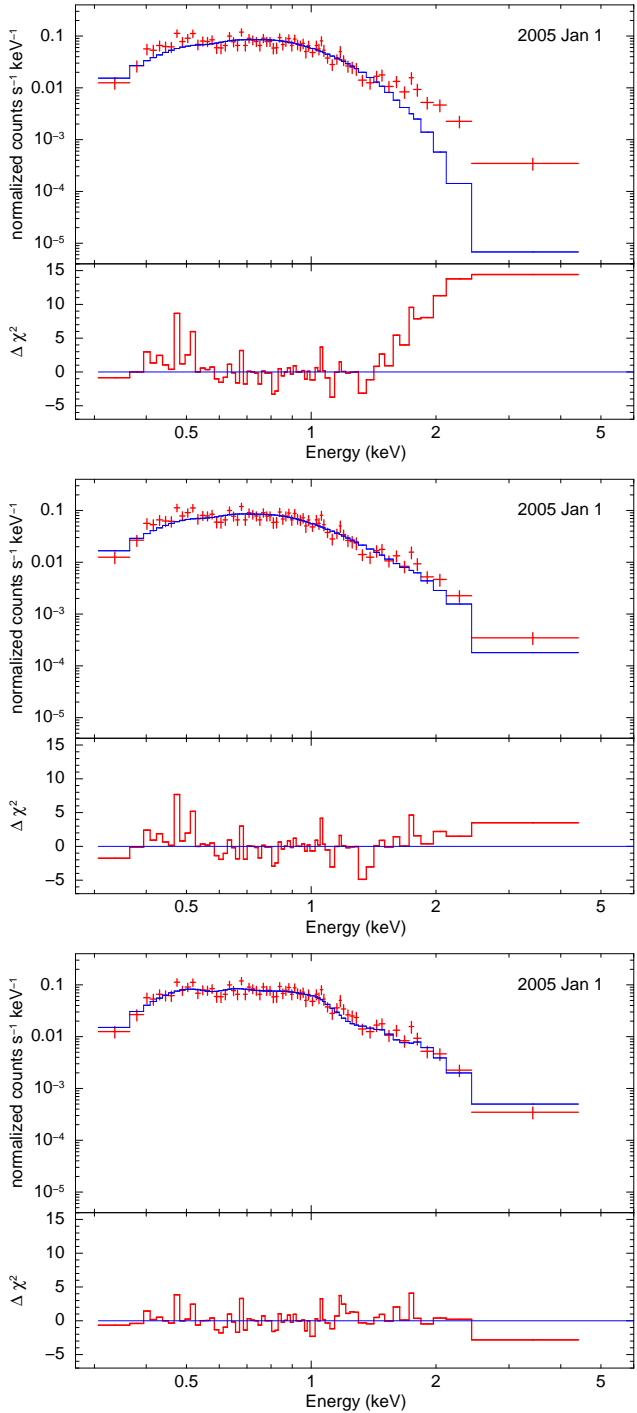

\begin{center}
\epsfig{figure=fig7a.ps,width=61mm,angle=270}\\[5pt]
\epsfig{figure=fig7b.ps,width=61mm,angle=270}\\[5pt]
\epsfig{figure=fig7c.ps,width=61mm,angle=270}
\end{center}
\caption{{\it {Chandra}}/ACIS-S spectral data and $\chi^2$ residuals 
for ObsID 4737.  Top panel: a simple blackbody does not provide 
an acceptable fit ($\chi^2_{\nu} > 2$). Middle panel: correcting for pile-up 
improves the fit, giving $\chi^2_{\nu} = 87.3/65$; however, systematic 
residuals are still clearly visible in the shape of the fitted spectrum.
Adding a single-temperature thermal-plasma component (not shown) 
improves the fit to $\chi^2_{\nu} = 79.8/63$. Bottom panel: blackbody 
model with pileup and two-temperature thermal plasma, giving 
$\chi^2_{\nu} = 63.5/61$. This is a significant improvement of the fit 
at the 99\% confidence level over a single-temperature thermal plasma 
model and over a model with no thermal plasma emission.  
See Table A4 for the value of the best-fitting parameters.
Datapoints have been binned to $> 15$ counts per bin before fitting.}
\label{fig7}
\end{figure}

\begin{figure}
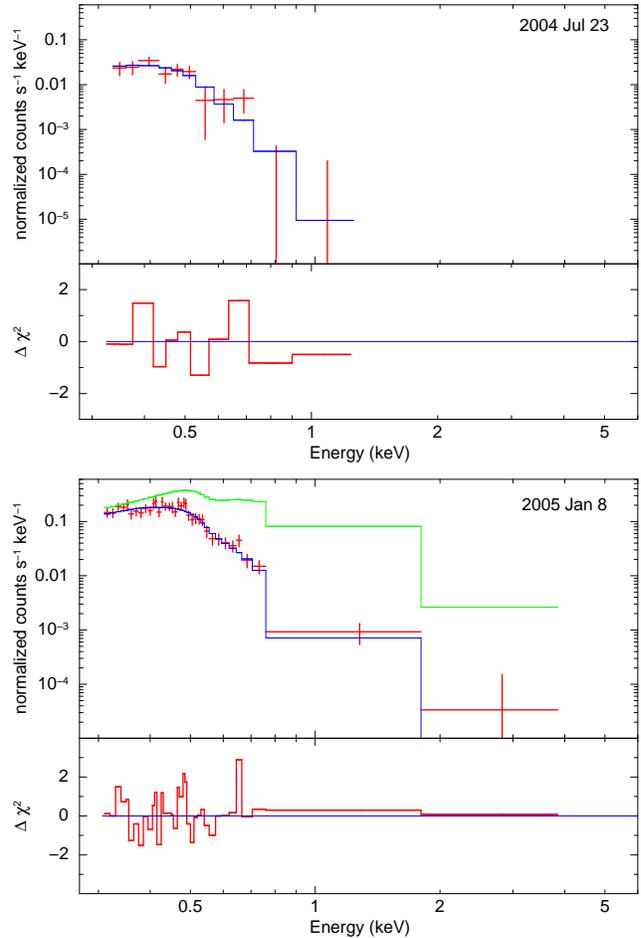

\begin{center}
\epsfig{figure=xmm_bb_plot_jul23.ps,width=61mm,angle=270}\\[5pt]
\epsfig{figure=xmm_bb_plot_jan8.ps,width=61mm,angle=270}
\end{center}
\caption{Top panel: {\it {XMM-Newton}}/EPIC spectral data and 
$\chi^2$ residuals for the 2004 July 23 observation, fitted with 
an absorbed blackbody model.
Bottom panel: {\it {XMM-Newton}}/EPIC spectral data and $\chi^2$ residuals 
for the 2005 January 8 observation, fitted with an absorbed blackbody 
model. The green histogram is the count rate that would have been observed 
in the same binned channels if the spectrum was identical to that 
observed in the {\it Chandra} ObsID 4737 ({\it i.e.}, one week earlier): 
the harder tail has completely disappeared. 
The best-fitting parameters for both observations are listed in Table A5.  
Datapoints have been binned to $> 20$ counts per bin before fitting.}
\label{fig8}
\end{figure}

\subsection{Hard component and edges}

A harder X-ray excess is detected in several epochs, with different properties.
The stacked dim state is the only X-ray spectrum dominated by hard emission (Figure 2, 
bottom right panel). 
Fitting it with a thermal component plus power-law, we obtain a photon index 
$\Gamma = 1.9^{+0.9}_{-0.7}$ and an unabsorbed 0.3--10 keV flux 
in the power-law component $f_{\rm pl} \approx 2 \times 10^{-15}$ 
erg cm$^{-2}$ s$^{-1}$, corresponding to an emitted power-law 
luminosity $L_{\rm pl} \approx 10^{37}$ erg s$^{-1}$.
The observed photon count rate due to the power-law component 
above 1 keV is only $\approx$$10^{-4}$ ct s$^{-1}$; 
the count rate above 1.5 keV is $\approx$$5 \times 10^{-5}$ ct s$^{-1}$.
There are not enough counts to attempt any physical 
interpretation of this component ({\it e.g.}, whether it is 
inverse-Compton emission, or synchrotron, or bremsstrahlung).
We re-fitted all the spectra taken in brighter epochs, adding a constant 
power-law component with the same slope and normalization determined 
in the dim state. We find that such faint component may have been present 
at all epochs but would not be significantly detected in the spectra of any of them. 
The 1.5--7 keV model count rate in the stacked dim state is also consistent 
with the very low hard-band count rate detected in the lightcurves at most epochs (Figure 2). 
Therefore, we cannot say that the hard component appeared or strengthened 
in the dim state (as is the case instead in X-ray binaries, 
when they move from the high/soft to the low/hard state): 
it could have been always present but be detectable only when 
the supersoft thermal component slipped out of the {\it Chandra} energy band, 
and only thanks to the very deep exposure time 
of the stacked dim-state observation (670 ks). 

A different (stronger) type of hard component is detected at some other 
epochs. This harder emission is most apparent in the two observations 
(2000 March 26 and 2005 January 1) taken when the source was brightest. 
We have already shown (Figures 1,2 and Table 2) that the harder emission 
has significant short-term variability, and hard-band flares 
often coincide with peaks in the soft-band count rate. 
Here, we investigate the harder emission further, focussing 
on its spectral properties.

The first possible explanation for the hard component is pile-up, 
as suggested by \cite{kong05}. 
From the observed count rate and fitted parameters of the soft component, 
we used {\small {PIMMS}} Version 4.7b{\footnote{http://cxc.harvard.edu/toolkit/pimms.jsp}}
to estimate that the average pile-up fraction is $\approx 10\%$ for ObsID 934, 
and $\approx 7\%$ for ObsID 4737. When we split ObsID 934 into three 
sub-intervals at high, medium and low count rates, 
we find a pile-up fraction as high as $\approx 15\%$ in the brightest  
interval, 
and negligible in the faintest one. Therefore, we know that pile-up is likely 
to affect the spectral appearance to some degree.
We convolved our blackbody model with pile-up models in {\small XSPEC} and 
{\small ISIS} and found that we can explain {\it some} of the hard emission 
in the bright state, as already shown by \cite{kong05}. However, 
we cannot explain all of it, and we still see systematic residuals 
in the spectral shape (top and middle panels of Figures 6 and 7 
for ObsID 934-high and ObsID 4737, respectively).

\begin{table}
\begin{center}
\begin{tabular}{lccr}\hline
\hline
Epoch & $R_{\rm bb}$ & $kT_{\rm bb}$ & $L^{\rm bol}_{\rm bb}$ \\[5pt]
 & ($10^3$ km) & (eV) & ($10^{39}$ erg s$^{-1}$)\\
\hline\\[-5pt]
2000 Mar 26 (H) & $10.2^{+0.4}_{-0.3}$ & $135^{+3}_{-4}$ & $4.6^{+0.2}_{-0.3}$\\[5pt]
2000 Mar 26 (M) & $10.5^{+0.3}_{-0.3}$ & $119^{+3}_{-3}$ & $2.9^{+0.2}_{-0.2}$\\[5pt]
2000 Mar 26 (L) & $18.4^{+0.9}_{-1.0}$ & $90^{+3}_{-3}$ & $2.9^{+0.3}_{-0.3}$\\[5pt]
2000 Oct 29 & $29.0^{+32.7}_{-14.4}$ & $77^{+11}_{-10}$ & $3.6^{+6.8}_{-2.0}$ \\[5pt]
2004 Jul 5--11 & $47.0^{+32.6}_{-17.0}$ & $69^{+7}_{-7}$ & $6.6^{+5.9}_{-2.9}$ \\[5pt]
2004 Jul 23  & $64.7^{+\ast}_{-54.1}$  & $48^{+20}_{-25}$ & $3.5^{+\ast}_{-3.3}$\\[5pt]
2004 Dec 22--24 & $>54$ & $<49$ & $> 2.1$ \\[5pt]
2004 Dec 30 & $43.5^{+29.3}_{-17.2}$ & $75^{+6}_{-6}$ & $7.7^{+8.6}_{-3.8}$ \\[5pt]
2005 Jan 1 & $22.5^{+10.3}_{-4.7}$ & $100^{+13}_{-10}$ & $6.6^{+1.2}_{-1.0}$ \\[5pt]
2005 Jan 8 & $101.5^{+82.7}_{-43.1}$ & $56^{+5}_{-5}$ & $12.8^{+17.9}_{-6.9}$ \\[5pt]
Faint state & $24.3^{+102.1}_{-10.8}$ & $53^{+8}_{-11}$ & $0.6^{+2.7}_{-0.2}$ \\
\hline
\end{tabular} 
\end{center}
\caption{Physical parameters of the supersoft thermal component in various epochs, 
fitted with a single-temperature blackbody model. Errors are 90\% confidence limits. 
See Tables A2, A4, A5 for the full set of spectral parameters in those fits.
The stacked ``faint state'' spectrum is defined as explained in Section 3; 
the 2000 March 26 observation is split into high (H), medium (M) and low (L) 
count-rate sub-intervals.}
\label{tab3}
\end{table}

\begin{table}
\begin{center}
\begin{tabular}{lccr}\hline
\hline
Epoch & $r_{\rm in} \, \sqrt{\cos \theta}$  & $kT_{\rm in}$ & $L^{\rm bol}_{\rm diskbb} \times \cos \theta$ \\[5pt]
 & ($10^3$ km) & (eV) & ($10^{39}$ erg s$^{-1}$)\\
\hline\\[-5pt]
2000 Mar 26 (H) & $5.9^{+0.3}_{-0.3}$ & $174^{+6}_{-7}$ & $4.1^{+0.2}_{-0.2}$\\[5pt]
2000 Mar 26 (M) & $4.3^{+0.2}_{-0.2}$ & $178^{+2}_{-2}$ & $2.4^{+0.2}_{-0.2}$\\[5pt]
2000 Mar 26 (L) & $13.0^{+1.1}_{-0.8}$ & $109^{+2}_{-2}$ & $3.1^{+0.4}_{-0.4}$\\[5pt]
2000 Oct 29 & $24.0^{+34.3}_{-12.3}$ & $90^{+14}_{-13}$ & $4.9^{+10.6}_{-2.8}$ \\[5pt]
2004 Jul 5--11 & $39.7^{+30.4}_{-16.5}$ & $81^{+9}_{-6}$ & $8.5^{+7.1}_{-4.0}$ \\[5pt]
2004 Jul 23 & $77.8^{+\ast}_{-70.3}$ & $52^{+28}_{-34}$ & $11.0^{+\ast}_{-10.5}$\\[5pt]
2004 Dec 22--24 & $>62$ & $<54$ & $>4.0$ \\[5pt]
2004 Dec 30 & $46.7^{+34.3}_{-12.3}$ & $84^{+8}_{-6}$ & $13.9^{+16.2}_{-7.5}$ \\[5pt]
2005 Jan 1 & $8.6^{+14.9}_{-3.5}$ & $147^{+6}_{-6}$ & $4.4^{+7.9}_{-1.1}$ \\[5pt] 
2005 Jan 8 & $103.7^{+103.5}_{-44.6}$ & $62^{+5}_{-5}$ & $20.5^{+35.4}_{-10.5}$ \\[5pt]  
Faint state & $24.5^{+128.3}_{-19.3}$ & $59^{+8}_{-6}$ & $1.0^{+16.6}_{-0.9}$ \\
\hline
\end{tabular} 
\end{center}
\caption{Physical parameters of the supersoft thermal component in the same epochs 
listed in Table 2, but fitted with a disk-blackbody model. Errors are 90\% confidence limits.}
\label{tab4}
\end{table}


For the high-count-rate interval of ObsID 934, 
the best-fitting piled-up blackbody model 
(Model 2 in Table A1, plotted in the middle panel of Figure 6) provides 
a still unsatisfactory $\chi^2_{\nu} = 128.7/78$. 
Adding a {\it mekal} component (Model 3 in Table A1) 
dramatically improves the fit, 
giving $\chi^2_{\nu} = 93.4/76$ 
(an F-test significance\footnote{\citet{protassov02} discuss caveats 
on the use of the F-test for the estimate of the significance 
of emission lines in X-ray spectra. We are aware of those caveats; however, 
we visually inspected our spectral fits with and without the additional 
absorption or emission features, and noticed clear systematic residuals 
when those components were not included.} 
at the 99.999\% level).
Adding a second {\it mekal} component
(Model 4 in Table A1, plotted in the bottom panel of Figure 6)
further improves the fit, giving $\chi^2_{\nu} = 84.9/74$: the F-test 
shows that this is a significant improvement at the 97\% level 
compared with the model with a single-temperature {\it mekal}. 
Adding a third {\it mekal} component to the fit does not lead 
to further significant improvements. However, when we fitted 
all three sub-intervals of ObsID 934 simultaneously (Table A2 and Figure 5), 
with locked {\it mekal} temperatures and column density, a third 
{\it mekal} component becomes significant in ObsID 934-high at the 98\% level.
The temperature range of the thermal-plasma components depends 
on whether we use a two-temperature or three-temperature approximation. 
Using a two-temperature {\it mekal} model, we find 
$kT_1 \approx 0.7$ keV and $kT_2 \approx 1.5$ keV; applying 
a three-temperature model, 
we find $kT_1 \approx 0.6$ keV, $kT_2 \approx 1$ keV and $kT_3 \ga 2$ keV.
An equivalent interpretation of this finding is that ObsID 934-high 
has an extended hard-tail emission consistent with multi-temperature 
thermal-plasma emission, from $kT \approx 0.6$ keV to $kT \approx 2$ keV. 
For the medium-count-rate interval of ObsID 934 (Figure 5, middle panel), 
the addition of a single {\it mekal} component at $kT \approx 0.6$ keV 
improves the piled-up blackbody fit with 99\% significance; 
a second {\it mekal} component does not lead to any further 
significant improvements. Instead, the fit is dramatically 
improved (F-test significance $>99.999\%$) by adding 
an absorption edge at $\approx$$1.07^{+0.03}_{-0.05}$ keV 
(Figure 5, middle panel).
Finally, ObsID 934-low (Figure 5, bottom panel) is well fitted 
with a simple (and cooler) blackbody spectrum, with no evidence 
of additional thermal plasma emission or edges.

Even a simple visual comparison of the spectral evolution between 
ObsID 934-high, ObsID 934-medium and ObsID934-low (top left panels 
of Figures 3,4, and Figure 5) strikingly confirms that the harder component 
is not just a marginal fitting residual or a pile-up artifact. 
The total unabsorbed 0.3--10 keV luminosity during ObsID 934-high 
is $L_{0.3-10} \approx 4.3 \times 10^{39}$ erg s$^{-1}$: of this, 
the contribution from the harder component is $\approx$7$ \times 10^{38}$ 
erg s$^{-1}$, the rest coming from the $\sim$0.1-keV optically thick 
blackbody emission. In ObsID 934-medium, the emission 
below $\approx$1 keV is approximately unchanged, but 
the harder flux has dramatically decreased; for example, 
an absorption edge appears to have replaced the higher-temperature 
thermal-plasma emission. In this regime, the thermal-plasma emission 
contributes just over $\approx$10\% of the unabsorbed luminosity 
in the 0.3--10 keV band, that is $\approx$3$ \times 10^{38}$ 
erg s$^{-1}$, compared with a total (supersoft blackbody plus harder 
components) 0.3--10 keV emitted luminosity of $\approx$2.3$ \times 10^{39}$ 
erg s$^{-1}$. Finally, ObsID 934-low is well fitted with a simple (and cooler) 
blackbody spectrum, with a 0.3--10 keV emitted luminosity 
of $\approx$1.5$ \times 10^{39}$ erg s$^{-1}$ and no evidence of additional 
harder components or edges.
In summary, within the single observation ObsID 934, the M\,101 ULS 
switched between a purely supersoft thermal state with 
$kT_{\rm bb} \approx 0.10$ keV, and a harder state 
with a broad-band tail detected at least up to 5 keV.

The other epoch when the M101 ULS spectrum displays a strong harder 
component is 2005 January 1 (ObsID 4737: bottom left panel of Figures 3,4).
Adding a single-temperature {\it mekal} component provides a better fit 
than a piled-up blackbody model, with 95\% significance; adding a second 
{\it mekal} component provides a further improvement with 99\% significance
(Figure 7). The best-fitting temperatures of the two components 
are $kT_1 \approx 0.7$ keV and $kT_2 \approx 1.3$ keV, consistent 
with the spread of temperatures modelled for Obs 934-high. 
The unabsorbed luminosity of the hard excess is $\approx$7$ \times 10^{38}$ 
erg s$^{-1}$, again similar to the value estimated for Obs 934-high.
Note that a week later, in the 2005 January 8 {\it XMM-Newton} observation, 
the hard tail had completely disappeared (Figure 8) and the blackbody component 
had cooled from $\approx$100 eV to $\approx$50 eV, while the fitted 
blackbody radius had increased from $\approx$20,000 km to $\approx$100,000 km.

ObsID 934 and ObsID 4737 have the highest observed 
count rates and highest 0.3--10 keV luminosities (Table 2), partly because 
the blackbody component is hotter in those two epochs, and partly because 
of their hard emission tails. Neither an excess hard component, 
nor an absorption edge is significantly detected in any other individual 
observation, as they are all consistent with a simple blackbody. 
However, we did find such features in the stacked 
spectrum from 2004 July 5--11 (Table 1 for the identification, 
and Table A4 for the fit parameters), because of the improved 
signal-to-noise ratio. A simple blackbody spectrum gives 
$\chi^2_{\nu} = 61.3/41$, while the fit improves to $\chi^2_{\nu} = 42.3/37$ 
with the addition of a {\it mekal} component at $kT \approx 0.6$ keV 
and an edge at $E \approx 0.93$ keV (left middle panel in Figures 3,4). 
Both the {\it mekal} component and the edge are significant 
at the 99\% confidence level. The unabsorbed luminosity 
of the hard excess is only $\approx$3$ \times 10^{37}$ erg s$^{-1}$, 
which is why it is only significantly detected in the stacked spectrum.
We summarize the energy and F-test significance of the excess emission and 
absorption components in Table 5.



\begin{table*}
\centering
\caption{Best-fitting parameters and F-test significance of additional spectral features (phenomenologically modelled as thermal-plasma emission or absorption edges) at all epochs in which they significantly improve the fit. Errors are 90\% confidence limits for single parameters.}
\begin{tabular}{lrrrrrr}
\hline\hline
\multicolumn{1}{c}{Epoch}&\multicolumn{3}{c}{Thermal Plasma}&\multicolumn{3}{c}{Edge}\\
\cline{2-7}\\[-6pt]
& $kT$ (keV) & Normalization & Significance & $E$ (keV) & $\tau_{\rm max}$ & Significance\\
\hline\\
2000 Mar 26 (H) & $0.61^{+0.06}_{-0.06}$ & 
$\left(1.7^{+1.0}_{-1.1}\right) \times 10^{-5}$ & $>99.9\%$ &&&\\[5pt]
& $0.98^{+0.16}_{-0.17}$ & $\left(3.9^{+1.4}_{-1.4}\right) \times 10^{-5}$ 
& $>95\%$ &&&\\[5pt]
& $2.5^{+*}_{-1.2}$ & $\left(1.9^{+1.9}_{-1.6}\right) \times 10^{-5}$ & 
$>95\%^1$ &&&\\[5pt]
2000 Mar 26 (M) & $0.61^{+0.06}_{-0.06}$ 
& $\left(2.6^{+0.5}_{-0.4}\right) \times 10^{-5}$ & $>99\%$ 
& $1.07^{+0.03}_{-0.03}$ & $2.1^{+1.3}_{-0.8}$ & $>99.9\%$\\[5pt]
2004 Jul 5--11 & $0.59^{+0.21}_{-0.26}$ 
& $\left(2.9^{+4.6}_{-1.5}\right) \times 10^{-6}$ & $>99\%$ 
& $0.93^{+0.05}_{-0.04}$ & $2.1^{+1.6}_{-0.9}$ & $>99\%$\\[5pt]
2005 Jan 1 & $0.70^{+0.17}_{-0.13}$ 
& $\left(3.1^{+1.5}_{-1.5}\right) \times 10^{-5}$ & $>95\%$ &&&\\[5pt]
 & $1.30^{+0.20}_{-0.20}$ 
& $\left(4.3^{+1.7}_{-1.6}\right) \times 10^{-5}$ & $>99\%$ &&&\\
\hline 
\end{tabular}
\begin{flushleft}
$^1$ This thermal plasma component is statistically equivalent to a {\it bremsstrahlung} or {\it comptt} component.
\end{flushleft}
\label{residual_tab}
\end{table*}

It is important to remark here that a {\it mekal} (or an {\it apec}) model 
is the simplest but not the only option for fitting the excess harder emission.
It provides a phenomenologically good tool at CCD resolution, 
with its moderately broad-band emission peaking around 1 keV 
and not contributing much below $\approx$0.5 keV and above $\approx$2 keV. 
However, it is not necessarily the correct physical interpretation.
For example, we verified that the non-equilibrium photo-ionization model 
{\it nei} also provides statistically equivalent fits. 
Thus, with the data at hand, we have no empirical elements 
to decide between collisionally ionized and photo-ionized gas.

Moreover, at moderately low signal-to-noise ratio and CCD spectral 
resolution, a phenomenological line-emission model can mimic 
a more physical model consisting of a smooth emission component 
above 1 keV (for example an inverse-Compton tail, or direct emission 
from the inner accretion disk) with a series 
of absorption edges (redshifted and blueshifted if the outflow is relativistic, 
as in the case of SS\,433) imprinted by the dense outflow 
\citep{fabrika07,middleton14}. 
In fact, the fast variability of the hard excess, appearing and disappearing 
sometime within a few 1000 s, is more likely to be caused by variable 
occultation of an inner compact source of hard X-ray photons 
emerging through a clumpy wind.
In this scenario, we speculate that the M\,101 ULS may have a pure 
supersoft blackbody spectrum when all the hard X-ray photons from 
the inner disk/corona are completely absorbed and reprocessed 
in an optically thick outflow; instead, it may show a harder tail at epochs  
when some of the hard X-ray emission emerges through the outflow 
(as is likely the case in the soft-ultraluminous regime of ULXs: 
\citealt{sutton13,middleton15}). The 2000 Mar 26 observation 
could be an example of a transition between a ULS regime 
(blackbody component at $kT_{\rm bb} \approx 0.10$ keV 
and no hard tail), and a two-component soft-ultraluminous ULX regime.
On the other hand, 
thermal plasma emission may also come from regions of the clumpy outflow 
that are just outside the photosphere, or when our line of sight is such 
that the outflow is not completely optically thick.

For ObsID 934-high, we tested whether an inverse-Compton component 
is consistent with the harder part of the emission tail. 
We replaced the higher temperature 
{\it mekal} component with a {\it comptt} component (Table A3), 
fixing its seed photon
temperature $kT_0$ equal to the blackbody temperature $kT_{\rm bb}$. 
The electron temperature $kT_e$ in the Comptonizing medium is not well 
constrained, because we do not have enough counts to detect 
a high-energy rollover; we arbitrarily fixed it at 4 keV, but any other 
choice of $kT_e > 2$ keV does not change the results of the fit. 
The scattering optical depth $\tau$ is $> 2.0$, 
but its upper limit is not constrained. Table A2 and A3 show  
that a Comptonization component provides as good a fit as a hot thermal 
plasma component. 

Regardless of the uncertainty on the physical origin 
of the hard excess, we stress that the main objective of this paper  
is to compare the characteristic radii, temperature and bolometric 
luminosities of the supersoft component with the predictions 
of a phenomenological outflow model. The fitted values of such quantities 
are robust and not strongly dependent on the first-order corrections 
due to residual harder components.

\section{Discussion}

\subsection{The case against the accretion disk model}
 
Both our timing and spectral results highlight severe problems 
for the disk interpretation. 
The strong short-term varibility is clearly inconsistent 
with the typical behaviour of standard accretion disks 
in accreting BHs. If there is an accretion disk in the M\,101 ULs, 
it is clearly not in a steady state. More likely, 
the system is undergoing rapid flaring, more typical of 
fast outflows than of thermal or viscous processes inside 
a standard disk.

For a standard-disk emission spectrum, there is a fundamental relation 
between the inner radius $R_{\rm in}$, the bolometric luminosity, 
and the peak temperature $T_{\rm in}$ \citep{ss73,makishima86,kubota98,frank02}. 
Assuming that the disk extends down to the innermost stable circular orbit, 
this provides a rough BH mass estimate, which is empirically satisfied within 
a factor of 2 for Galactic stellar-mass BHs: 
\begin{eqnarray}
\nonumber
M & \approx & 10.0 \, \left(\frac{\eta}{0.1}\right) \, 
              \left(\frac{\xi \kappa^2}{1.19}\right) \,
              \left(\frac{L^{\rm bol}_{\rm diskbb}}{5 \times 10^{38} \rm{~erg~s}^{-1}}\right)^{1/2} 
     \\ 
           &\times&   \left(\frac{kT_{\rm in}}{1\rm{~keV}}\right)^{-2} \, M_{\odot}
\end{eqnarray}
\citep{soria07}, where $\eta$ is the radiative efficiency 
(in a semi-classical approximation, $\eta \approx 1/12$ 
for a Schwarzschild BH, and $\eta \approx 1/2$ for a maximally spinning 
Kerr BH), $\kappa$ is a spectral hardening factor 
(ratio between colour temperature and effective temperature: 
\citealt{shimura95}), 
$\xi$ is a normalization factor introduced because the peak 
temperature occurs outside the apparent inner radius \citep{kubota98}. 
For the M\,101 ULS, the disk-blackbody temperatures and luminosities 
of the thermal component 
(Table 4) would require a BH mass of a few times $10^3 M_{\odot}$ accreting at 
$\sim$1\% of its Eddington limit. This is inconsistent with 
the results of \cite{liu13}. Moreover, it is not self-consistent 
even within an IMBH scenario, because we do not expect 
an accreting BH to be in the high/soft state at such low Eddington 
ratios (see Equation 1), without a harder power-law component.

Finally, it is clear that the fitted value of $R_{\rm in}$ 
in the various observations is not consistent with being constant 
(Table 3), as should be for a disk in the high/soft state. 
This is equivalent to saying that there is also no 
$L^{\rm bol}_{\rm diskbb} \propto T^4_{\rm in}$ trend  
in the fitted distribution of temperatures and luminosities.
There is instead an anticorrelation $R \propto T^{-2}$ between fitted radius 
and temperatures (either in the blackbody or disk-blackbody model), 
which is more consistent with an expanding or contracting photosphere.

We are aware that none of those arguments is conclusive 
on its own. There are BHs in the high/soft state BHs that 
do not follow the $L^{\rm bol}_{\rm diskbb} \propto T^4_{\rm in}$ 
relation ({\it e.g.}, LMC X-1: \citealt{gierlinski04}).
This may be due to changes in the disk atmosphere, causing 
changes in the hardening factor \citep{salvesen13,reynolds13, 
walton13}, instead of changes in the innermost radius of the disk. 
Nonetheless, when taken together, the lack of a standard disk track, 
the low luminosity for a high/soft state, and the high short-term variability 
make the standard disk interpretation very problematic for the M\,101 ULS. 
It was also shown \citep{urquhart15} that this source  
is not a lone exception: several other ULSs share similar properties, 
suggesting that the standard disk interpretation is not viable 
for the whole ULS class.


\subsection{The outflow photosphere model}

Given the serious problems of the standard disk interpretation 
(both in the stellar-mass and IMBH scenarios), we now examine 
the alternative possibility that the thermal emission in ULX-1 
and other supersoft ULXs comes 
from the photosphere of a radiatively-driven outflow launched 
from an accretion disk in the super-Eddington accretion regime.
Here, we follow and then extend the analysis of \cite{shen15}, 
and we refer the readers to that paper for more detailed discussion 
of the outflow parameters. 

\cite{shen15} define and solve a system of five equations for the outflow. 
Firstly, an equation for the absorption opacity $\kappa^a_\nu$:
\begin{equation}
\kappa^a_\nu = C \rho \, T^{-7/2} \mathrm{~cm}^2  \mathrm{~g}^{-1}
\end{equation}
where $\rho$ is the outflow gas density and $C \approx 2.4 \times 10^{25}$.
The Thomson electron scattering opacity is $\kappa_{\rm s} \approx 0.2(1+X)$ cm$^{-2}$ g$^{-1}$, 
where $X$ is the hydrogen mass fraction. Henceforth, we assume that 
scattering dominates over absorption (as is the case in the inner 
part of accretion flows onto stellar-mass BHs), so that 
$\kappa_{\rm s} \gg \kappa^a_\nu$. 
The second equation defines the thermalization radius $R_{\rm th}$ 
where the {\it effective} absorption optical depth $\tau^\ast_\nu \equiv 1$:
\begin{equation}
\tau^\ast_\nu(R_{\rm th}) = \int_{R_{\rm th}}^{\infty} \rho 
                \sqrt{\kappa^a_\nu \left(\kappa^a_\nu + \kappa_{\rm s}\right)} \, {\mathrm{d}}r 
                \approx \int_{R_{\rm th}}^{\infty} \rho  \sqrt{\kappa^a_\nu \kappa_{\rm s}} \, {\mathrm{d}}r
                 = 1.
\end{equation}
We are assuming that the outflow is accelerating 
(driven by radiation pressure) before reaching a constant speed: 
therefore, the outflow density drops at least as fast as $r^{-2}$. 
This simplifies the integration in Equation (4) and gives \citep{shen15}:
\begin{equation}
\rho(R_{\rm th}) R_{\rm th} \sqrt{\kappa^a_\nu \kappa_{\rm s}} \approx 1.
\end{equation}

Inside the thermalization radius, photons are in thermal equilibrium 
with the gas; beyond $R_{\rm th}$, they are decoupled, in the sense 
that they are no longer absorbed and re-emitted (they can still scatter 
multiple times). 
Another characteristic radius of the outflow is the photon trapping radius $R_{\rm tr}$: 
inside $R_{\rm tr}$, photons are advected with the flow (diffusion timescale 
longer than the expansion timescale).
As discussed by \cite{shen15}, if the gas outflow is such that $R_{\rm th} > R_{\rm tr}$, 
the colour temperature of the photon spectrum observed at infinity is the temperature 
at the thermalization radius: $T_{\rm th} \equiv T(R_{\rm th}) = T_{\rm bb}$ 
\citep{shen15}. If $R_{\rm th} < R_{\rm tr}$, instead, the observed temperature 
$T_{\rm bb} = T(R_{\rm tr})$.
Here, we follow the derivation for the case of the thermalization radius 
larger than the trapping radius: we will check later that this is indeed the case 
for the range of parameters suitable to the M\,101 ULS.

The third equation in the system of \cite{shen15} is the luminosity 
density from the radiative diffusion equation, which can be integrated 
over all photon energies to give:
\begin{equation}
L \approx \frac{16}{3} \pi R_{\rm th}^2 \frac{\sigma T_{\rm th}^4}{\tau_\nu(R_{\rm th})},
\end{equation}
where $\tau_\nu(R_{\rm th})$ is the {\it total} optical depth (in our case, dominated 
by scattering), that is 
\begin{equation}
\tau_\nu(R_{\rm th}) = \int_{R_{\rm th}}^{\infty} \rho \left(\kappa^a_\nu + \kappa_{\rm s}\right) {\mathrm{d}}r 
   \approx \rho(R_{\rm th}) \kappa_{\rm s} R_{\rm th} = \tau_{\rm s}(R_{\rm th}).
\end{equation}

Finally, the fitted blackbody radius of the emerging radiation is {\it defined} as 
\begin{equation}
R_{\rm bb} \equiv \left(\frac{L}{4\pi \sigma T_{\rm bb}^4} \right)^{1/2}
\end{equation}
(notice that $R_{\rm bb} \neq R_{\rm th}$ even in the approximation that $T_{\rm bb} = T_{\rm th}$). 
By solving the system of equations (3), (5), (6), (7) and (8), \cite{shen15} 
determine the values of $R_{\rm th}$, $\rho(R_{\rm th})$ and $\tau_{\rm s}(R_{\rm th})$ 
as a function of the observable quantities $R_{\rm bb}$, $T_{\rm bb}$ and $L$.

We want to go a step further, and express $R_{\rm bb}$, $T_{\rm bb}$ and $L$ 
as a function of BH mass $M \equiv m M_{\odot}$ and mass accretion rate $\dot{M}$ at infinity, 
\begin{equation}
\dot{M} \equiv \dot{m} \, \dot{M}_{\rm Edd} 
           \approx \frac{2.5 \times 10^{39}}{(1+X)\, c^2} \, m \, \dot{m} {\mathrm{~g~s}}^{-1}, 
\end{equation}
where we have defined the Eddington luminosity 
\begin{equation}
L_{\rm Edd} \equiv 0.1 \dot{M}_{\rm Edd} \, c^2 \approx \frac{2.5 \times 10^{38}}{(1+X)} \, m {\mathrm{~erg~s}}^{-1} 
\end{equation}
for a hydrogen mass fraction $X$.
Note that \cite{shen15} take $L_{\rm Edd} \approx 2 \times 10^{38} m$ erg s$^{-1}$ 
because they assume a Wolf-Rayet donor star (following \cite{liu13}) and 
therefore hydrogen-poor accretion flow; for the same reason, they take 
$\kappa_{\rm s} \approx 0.2$ cm$^{-2}$ g$^{-1}$.  
For our purpose, we need to introduce two additional equations 
that express for example $\rho(R_{\rm th})$ and $L$ as a function of $m$, $\dot{m}$.
This was elegantly done in the outflow model of \cite{poutanen07}, 
and we shall follow their lead (see also similar treatments 
in \citealt{strubbe09,lodato11}).

Let us start from the luminosity. The power output $L_0$ 
of an outflow-dominated, super-Eddington accreting disk is 
$L_0 = L_{\rm Edd} \left(1 + \frac{3}{5} \ln \dot{m} \right)$.
However, a fraction $\epsilon_{\rm w}$ of that power is spent to accelerate the outflow; 
only a fraction $(1-\epsilon_{\rm w})$ emerges 
as radiative luminosity. 
Hence, the observed bolometric luminosity $L$ is:
\begin{equation}
L = (1-\epsilon_{\rm w}) \, L_{\rm Edd} \left(1 + \frac{3}{5} \ln \dot{m} \right) 
\end{equation}
with $\epsilon_{\rm w} \sim 0.5$ \citep{lipunova99,poutanen07}.
The amount of power used for accelerating the outflow depends of course 
on the mass loss rate in the outflow. That is a fraction $f_{\rm out}$ 
of the mass accretion rate at infinity: $\dot{M}_{\rm w} \equiv f_{\rm out} \dot{M}$.
The kinetic fraction $\epsilon_{\rm w}$ is related to the 
mass outflow fraction $f_{\rm out}$ by the useful approximation 
$f_{\rm out} \approx 0.83 \epsilon_{\rm w} - 0.25 \epsilon^2_{\rm w} $ 
\citep{poutanen07}.
For example, $f_{\rm out} = 0.5$ corresponds to $\epsilon_{\rm w} \approx 0.79$.

Our second additional equation relates the outflow density 
to the accretion rate: for the conservation of mass, 
\begin{equation}
\rho(R_{\rm th}) = \frac{\dot{M}_{\rm w}}{4\pi R_{\rm th}^2 v_{\rm w}} = \left(\frac{f_{\rm out}}{f_{\rm v}} \right)
      \, \frac{\dot{M}}{4\pi R_{\rm th}^2 v_{\rm esc}},
\end{equation}
where we have plausibly assumed that the outflow speed near the photosphere 
is of order of the escape velocity at the launching radius, 
that is $v_{\rm w} \equiv f_{\rm v} v_{\rm esc}$. For the outflow launching 
radius, we assume that most of the wind comes from around the spherization radius 
$R_{\rm sp} \propto \dot{m}$ \citep{king03,ss73}. \cite{poutanen07} show that  
\begin{equation}
R_{\rm sp}/R_{\rm in} \approx \left[1.34 - 0.4 \epsilon_{\rm w} + 0.1 \epsilon_{\rm w}^2\right] \dot{m} 
 - (1.1-0.7\epsilon_{\rm w}) \dot{m}^{1/3}.
\end{equation}
Considering the uncertainty of the observed quantities and 
the other approximations of the model, 
we avoid unnecessary complications and take 
\begin{equation}
R_{\rm sp} \approx 1.1 \dot{m} R_{\rm in}
\end{equation}
where $R_{\rm in} = (6GM/c^2)$ for a non-rotating BH.
Then, 
\begin{equation}
v_{\rm esc} = \left(\frac{2GM}{R_{\rm sp}}\right)^{1/2} \approx 0.55 \frac{c}{\sqrt{\dot{m}}}.
\end{equation}
Substituting Equations (9) and (15) into Equation (12):
\begin{eqnarray}
\rho(R_{\rm th}) &=& \left(\frac{0.83 \epsilon_{\rm w} - 0.25 \epsilon^2_{\rm w}}{f_{\rm v}} \right)\, 
          \frac{\dot{M}}{4\pi R_{\rm th}^2} \, \frac{\sqrt{\dot{m}}}{0.55c} \nonumber\\
           &= & \left(\frac{0.83 \epsilon_{\rm w} - 0.25 \epsilon^2_{\rm w}}{f_{\rm v}} \right)\, 
                 \frac{4.6 \times 10^{39}}{(1+X)\, c^3} \,\frac{m\, \dot{m}^{3/2}}{4\pi R_{\rm th}^2}.
\end{eqnarray}
Equations (11) and (16), added to the set of equations in \cite{shen15}, 
allow us to model the observed temperature and luminosity as a function 
of BH mass and accretion rate.

From Equations (6), (7) and (8), 
\begin{equation}
R_{\rm th} = \sqrt{\frac{3}{4}} \, \sqrt{\tau_{\rm s}} \, R_{\rm bb},
\end{equation}
\begin{equation}
\rho(R_{\rm th}) = \frac{\sqrt{\tau_{\rm s}}}{\sqrt{(3/4)} \, \kappa_{\rm s} \, R_{\rm bb}}.
\end{equation}
Substituting Equations (3), (6), (8), (17) and (18) into Equation (5):
\begin{equation}
\tau_{\rm s}(R_{\rm th}) = \left(\frac{3T_{\rm bb}L\kappa_{\rm s}^4}{16 \pi \sigma C^2} \right)^{1/5}.
\end{equation}
Moreover, from Equations (16) and (17):
\begin{equation}
\rho(R_{\rm th}) = \left(\frac{0.83 \epsilon_{\rm w} - 0.25 \epsilon^2_{\rm w}}{f_{\rm v}} \right)\, 
                 \frac{6.1 \times 10^{39}}{(1+X)\, c^3} \,\frac{m\, \dot{m}^{3/2}}{4\pi \tau_{\rm s} R_{\rm bb}^2}.
\end{equation}
Dividing Equation (20) by Equation (18):
\begin{equation}
R_{\rm bb} = 3.60 \times 10^6 \left(\frac{0.83 \epsilon_{\rm w} - 0.25 \epsilon^2_{\rm w}}{f_{\rm v}} \right)\,
       \tau_{\rm s}^{-3/2} \, m\, \dot{m}^{3/2} \mathrm{~cm}, 
\end{equation}
and from Equations (8) and (11):   
\begin{eqnarray}
T_{\rm bb} &=& \frac{1.28 \times 10^7}{(1+X)^{1/4}} \, 
        \frac{f_{\rm v}^{1/2} \, \left(1-\epsilon_{\rm w}\right)^{1/4}}
        {\left(0.83 \epsilon_{\rm w} - 0.25 \epsilon^2_{\rm w}\right)^{1/2}} 
        \, \tau_{\rm s}^{3/4} \nonumber \\
        & \times & m^{-1/4} \, \dot{m}^{-3/4} \left(1 + \frac{3}{5} \ln \dot{m} \right)^{1/4} 
        {\mathrm{~K}}.
\end{eqnarray}
We solve for the scattering optical depth $\tau_{\rm s}(m, \dot{m})$ by inserting Equations (11) and (22) 
into (19):
\begin{eqnarray}
\tau_{\rm s}(R_{\rm th}) &=& 2189 \, \left( \frac{0.83\epsilon_{\rm w} - 0.25 \epsilon^2_{\rm w}}{f_{\rm v}} \right)^{-6/11} \, 
    \left(1-\epsilon_{\rm w}\right)^{7/11} \nonumber\\ 
     &\times& (1+X)^{9/11} \, m^{1/11} \, \dot{m}^{-9/11} \nonumber\\ 
     &\times& \left(1 + \frac{3}{5} \ln \dot{m} \right)^{7/11}.
\end{eqnarray}
Finally, we re-insert $\tau_{\rm s}(m, \dot{m})$ into Equations (21) and (22).
After long but straightforward algebra, we obtain our final equations for the observed 
blackbody radius and temperature as a function of BH mass and accretion rate:
\begin{eqnarray}
R_{\rm bb} &=& 35.2 \, \left( \frac{0.83\epsilon_{\rm w} - 0.25 \epsilon^2_{\rm w}}{f_{\rm v}} \right)^{20/11} \, 
    \left(1-\epsilon_{\rm w}\right)^{-21/22} \nonumber\\  
     &\times& (1+X)^{-27/22} \, m^{19/22} \, \dot{m}^{30/11} \nonumber\\
     &\times& \left(1 + \frac{3}{5} \ln \dot{m} \right)^{-21/22} \mathrm{~cm}, 
\end{eqnarray}
\begin{eqnarray}
T_{\rm bb} &=& 4.10 \times 10^9 \, \left( \frac{f_{\rm v}}{0.83\epsilon_{\rm w} - 0.25 \epsilon^2_{\rm w}} \right)^{10/11} \, 
    \left(1-\epsilon_{\rm w}\right)^{8/11} \nonumber\\  
     &\times& (1+X)^{4/11} \, m^{-2/11} \, \dot{m}^{-15/11} \nonumber\\ 
     &\times& \left(1 + \frac{3}{5} \ln \dot{m} \right)^{8/11} \mathrm{~K}, 
\end{eqnarray}
approximately scaling as $R_{\rm bb} \propto T_{\rm bb}^{-2}$ as expected.

\begin{figure*}
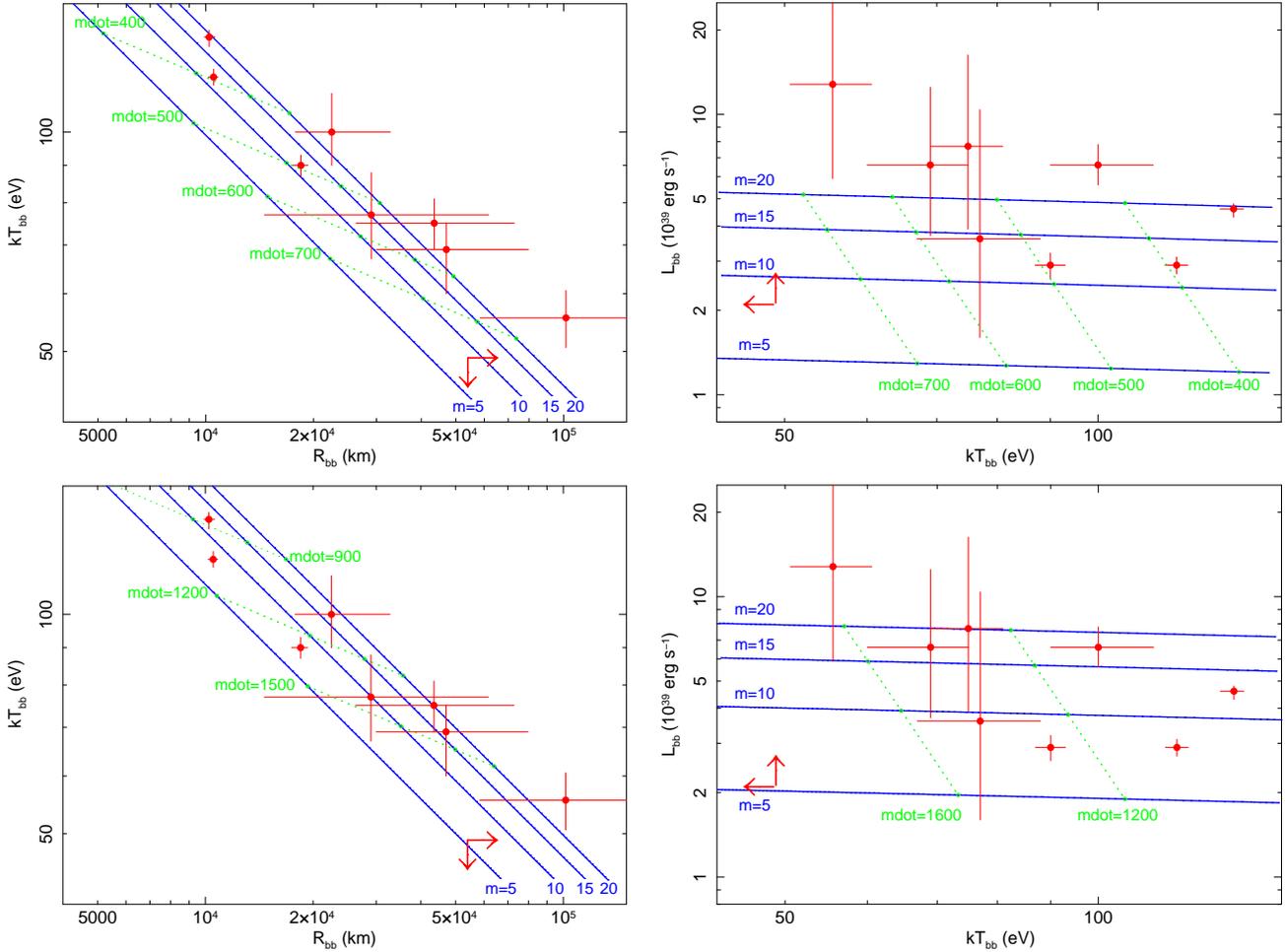

\begin{center}
\epsfig{figure=fig9a.ps,width=63mm,angle=270}
\hspace{0.2cm}
\epsfig{figure=fig9b.ps,width=63mm,angle=270}\\[5pt]
\epsfig{figure=fig9c.ps,width=63mm,angle=270}
\hspace{0.2cm}
\epsfig{figure=fig9d.ps,width=63mm,angle=270}
\end{center}
\caption{Predicted physical parameters from our super-Eddington outflow model, 
compared with the observations. 
Each solid blue curve represents the predicted values for a fixed BH mass 
(labelled next to each curve) over a range of accretion rates $\dot{m}$. 
Representative values of the accretion rates along each curve are also plotted 
(dotted green curves). The fitted datapoints from the {\it Chandra} observations 
are plotted in red (blackbody model: Table 3).
Top left: colour temperature versus blackbody radius, assuming hydrogen-poor 
accretion ($X=0$), $\epsilon_{\rm w} = 0.79$, $f_{\rm v} = 1$. 
Top right: bolometric luminosity versus colour temperature, for $X=0$, 
$\epsilon_{\rm w} = 0.79$, $f_{\rm v} = 1$.
Bottom left: colour temperature versus blackbody radius, assuming solar 
abundance ($X=0.73$), $\epsilon_{\rm w} = 0.5$, $f_{\rm v} = 1$. 
Bottom right: bolometric luminosity versus colour temperature, for 
$X=0.73$, $\epsilon_{\rm w} = 0.5$, $f_{\rm v} = 1$.}
\label{fig9}
\end{figure*}

\subsection{Comparing the outflow model with the data}

We now check whether our outflow photosphere model {\it can} produce 
blackbody temperatures, radii and luminosities consistent 
with the observed values (Table 3), for reasonable 
choices of parameters. We assume $f_{\rm v} = 1$, $f_{\rm out} = 0.5$ 
({\it i.e.}, half of the accretion inflow is re-ejected in the wind 
before reaching the BH), $X=0$, and a range of BH masses between 
$5M_{\odot}$ and $20 M_{\odot}$.
We find that for $\dot{m} \approx 400$--700, the model does indeed 
predict colour temperatures $\sim$50--130 eV and characteristic radii 
$\sim$10,000--100,000 km, for BH masses $\approx$10--20$M_{\odot}$ 
(Figure 9, top left). Moreover, the bolometric luminosity 
predicted for the same range of BH masses is 
$L^{\rm bol}_{\rm bb} \approx 2$--$5 \times 10^{39}$ erg s$^{-1}$ 
(Figure 9, top right), 
almost independent of accretion rate (and therefore of temperature 
and radius), in agreement with the observations. (Note again the absence 
of any $L \propto T^4$ trend in the observational data.)
Different choices of parameters within a plausible range can produce
other model runs in agreement with the data: for example, 
for $X=0.73$ (solar abundance) and $\epsilon_{\rm w} = 0.5$ 
(corresponding to $f_{\rm out} \approx 1/3$), the fitted radii, 
temperatures and luminosities are obtained for BH masses 
$\approx$10$M_{\odot}$ and accretion rates $\dot{m} \approx 900$--1600 
(Figure 9, bottom panels).
 
We can also verify a posteriori that we were justified in assuming 
$R_{\rm th} > R_{\rm tr}$. For this condition to occur, the outflow velocity 
has to be smaller than a critical trapping speed \citep{shen15}:
\begin{equation}
v_{\rm w}  \approx 0.55 \, f_{\rm v} \frac{c}{\sqrt{\dot{m}}} < v_{\rm crit} \equiv \frac{c}{\tau_\nu} 
    \approx \frac{c}{\tau_{\rm s}}.
\end{equation}
For $X=0$ and $f_{\rm out} = 0.5$, we showed that 
the observational data suggests $\dot{m} \approx 400$--700. 
The characteristic outflow speed is $v_{\rm w} \approx 6000$--8000 km s$^{-1}$, 
while $\tau_{\rm s} \approx 20$--30 (from Equation 23), and therefore 
$v_{\rm crit} \approx 10000$--15000 km s$^{-1}$.
For $X=0$ and $f_{\rm out} \approx 1/3$, $\dot{m} \approx 1200$--1600, 
implying a characteristic outflow speed $v_{\rm w} \approx 4000$--6000 km s$^{-1}$.
In this case, $\tau_{\rm s} \approx 30$--40, and therefore 
$v_{\rm crit} \approx 7500$--10000 km s$^{-1}$.

The kinetic power of the outflow at large distances can be estimated as follows. Let us assume
(as representative values) an asymptotic outflow speed $v_{\rm w} \approx 7000$ km s$^{-1}$, 
$\dot{m} \approx 500$, and because $f_{\rm out} = 0.5$, $\dot{m}_{\rm w} \approx 250$. 
For a $10$-$M_{\odot}$ BH, this corresponds to an outflow rate 
$\dot{M}_{\rm w} \approx 4 \times 10^{21}$ g s$^{-1}$, and a kinetic power $\approx 10^{39}$ 
erg s$^{-1}$. The photon luminosity emitted at the base of the outflow 
is $L = L_{\rm Edd} [1 + (3/5) \ln \dot{m}] \approx 10^{40}$ erg s$^{-1}$ 
(larger than the photon luminosity at the photosphere by a factor $1/(1-\epsilon_{\rm w})$).
A mechanical power in the fast wind $P_{\rm w} \approx 10\%$ of the (initial) 
photon luminosity is similar to what has been observed in fast-accreting AGN 
({\it e.g.}, \citealt{tombesi15}) and predicted in MDH simulations of super-Eddington 
accretion ({\it e.g.}, \citealt{jiang14}).
It is interesting to note that the most luminous standard ULXs within distances 
$\la 10$ Mpc reach X-ray luminosities of a few $\times 10^{40}$ erg s$^{-1}$ 
\citep{swartz04,walton11,swartz11}, while the M\,101 ULS 
(and other ULSs: \citealt{urquhart15}) 
reach luminosities of only a few $\times 10^{39}$ erg s$^{-1}$. 
We speculate that the upper envelope of total power output in ULSs  
corresponds to the upper envelope of luminosities in standard ULXs, 
after accounting for the reprocessing of a larger fraction of 
X-ray photons in the thicker outflow.

Another self-consistency check is required before we can use 
this outflow model. 
The thermalization radius must be larger than the launching radius (Equations 
17 and 14, respectively). (Note that the physical thermalization radius 
is $\approx$4--5 times larger than the ``apparent'' blackbody radius 
derived from spectral fitting: see Equation 17). 
Substituting into the solution for $R_{\rm th}$, this requires: 
\begin{eqnarray}
\frac{R_{\rm th}}{R_{\rm sp}} &=& 1.5 \times 10^{-3} \, 
    \left( \frac{0.83\epsilon_{\rm w} - 0.25 \epsilon^2_{\rm w}}{f_{\rm v}} \right)^{17/11} \, 
    (1+X)^{9/11}   \nonumber\\  
     &\times& \left(1-\epsilon_{\rm w}\right)^{-7/11}  \, m^{-1/11} \, \dot{m}^{29/22} \nonumber\\
     &\times& \left(1 + \frac{3}{5} \ln \dot{m} \right)^{-7/11} > 1.
\end{eqnarray}
For the first set of parameters discussed above ($X=0$ and $f_{\rm out} = 0.5$), 
this condition is satisfied for $\dot{m} \ga 380$. For the alternative set of parameters 
(solar abundance and $f_{\rm out} \approx 1/3$), it is satisfied 
only for $\dot{m} \ga 1250$. 
However, our calculations were done assuming a spherically symmetric outflow 
and a Schwarzschild BH 
in the definition of spherization and launching radius. 
In practice, the outflow will be thicker closer to the equatorial plane; therefore 
a lower $\dot{m}$ will be required to create an optically thick photosphere viewed 
from high inclination angles. As for the effect of BH spin,  
it is straightforward to solve the equations again 
for $R_{\rm in} = \alpha GM/c^2$ with $1< \alpha < 6$, and 
verify that for fixed $\dot{m}$ and decreasing $\alpha$, 
$R_{\rm th}/R_{\rm sp}$ scales approximately as $1/\sqrt\alpha$, 
and condition (27) is easier to satisfy. 

The general physical significance of this model is that 
for any given line of sight, as the mass inflow rate and, as a result, 
also the outflow rate increase, at some point the wind will become 
effectively optically thick (as opposed to just optically thick 
to scattering) and will develop a thermalization surface. 
When that happens, the inner region of the inflow is shrouded 
by a photosphere (effective optical depth $> 1$) which completely blocks 
our view of the inner disk/corona structure and makes the system look 
like a simple blackbody emitter with no high-energy tail. 
We used a spherical approximation for the wind in our model. 
We are aware that in reality, the outflow cannot be spherical, because 
it is launched from the accretion disk (among other reasons): 
for a fixed $\dot{m}$, the thickness of the wind will be higher 
for high-inclination line-of-sights, and lower when a source is seen 
more face-on, or even down the polar funnel, where the outflow is very 
tenuous and radiation escapes freely. Conversely, the higher 
the inclination, the lower the value of $\dot{m}$ required 
to make the wind effectively optically thick.
Nonetheless, we believe that a simple spherical approximation is useful 
to highlight the physical concept (in the same sense that a spherical 
approximation is used to define the Eddington accretion rate), 
and to estimate at least a characteristic 
order of magnitude for the accretion rate at which the wind at intermediate 
viewing angles is likely to become effectively optically thick.
Besides, for accretion rates as high as those discussed here ($\dot{m} \sim$ 
a few 100), the polar funnel is predicted to be quite narrow and 
the outflow is likely to cover most of the $4\pi$ solid angle around the BH. 
For example, it was proposed by \citet{king09} that the half-opening-angle 
$\theta$ of the polar outflow scales as $\theta \sim (150/\dot{m}^2)^{0.5}$.
The viewing-angle dependence on a ULX apprearance was already extensively 
discussed in \citet{sutton13} and \citet{middleton15}. 
The spectra of ObsID 934-high and ObsID 4737 might be a transitional stage 
between the supersoft regime and the standard ULX regime 
(soft thermal component plus high-energy tail carrying most of the flux), 
as the photospheric radius seen along our line of sight shrinks to the point 
where we start getting a direct view of harder photons emitted closer to the BH.
Photospheric temperatures $kT \approx 0.13$ keV may be the threshold 
between the two regimes.


\subsection{Donor star and binary period}

An important feature of our spectral fits, reproduced by the outflow model, 
is that the bolometric luminosity is $\ga 10^{39}$ erg s$^{-1}$ even in the X-ray 
faint states. If accretion is radiatively efficient, this requires 
a long-term-average accretion rate $\sim$ a few $10^{-7} M_{\odot}$ yr$^{-1}$, 
which is probably already too high to be consistent with wind accretion, advocated 
by the \cite{liu13} model. In fact, in our photosphere model, accretion 
is highly super-Eddington ($\dot{m} \sim$ a few 100) 
and therefore less efficient. If the radiative 
efficiency of super-Eddington accretion flows scales as 
$\eta \sim 0.1\,(1+\ln \dot{m})/\dot{m}$, the required mass 
accretion rate at infinity is $\approx$$3 \times 10^{-4} M_{\odot}$ yr$^{-1}$.
The BH SS\,433 is the only system in our Galaxy known so far that reaches 
such huge accretion rates \citep{king99,fabrika04,begelman06,king09}. 
Such mass tranfer rates can occur over the thermal timescale or nuclear timescale 
of massive donor stars in their late stages of evolution \citep{rappaport05}.
In particular, \cite{wiktorowicz15} showed that a 10-$M_{\odot}$ BH can 
be fed by a massive donor ($M_2 \approx 10 M_{\odot}$) in the short-lived 
Hertzsprung gap phase, reaching a Roche-lobe overflow rate 
of $\sim 10^{-3} M_{\odot}$ yr$^{-1}$ ($\dot{m} \sim 1000$). 

Crucially, such high accretion rates require that the donor star fills 
its Roche lobe. This is not the case for the orbital parameters 
measured by \cite{liu13}. If their claimed orbital period 
of $\approx (8.2 \pm 0.1)$ days (based on radial velocity measurements 
of the He {\footnotesize{II}} $\lambda$4686 emission line)
is confirmed by future observations, our super-Eddington outflow model is all 
but ruled out. Conversely, we suggest that the empirical measurement 
of the orbital period may be incorrect, and the true period may be 
$\la$3 days, so that the donor star does fill its Roche lobe.

One reason why we are not entirely convinced about the value of the period 
claimed by \cite{liu13} is that the He {\footnotesize{II}} emission 
was assumed to originate 
mainly from the Wolf-Rayet secondary with little contribution 
from the accretion disk. However, we do not know 
whether, or how much, the observed emission line is
contaminated by emission from the hot gas inside the BH 
Roche lobe (outer disk, outflow, accretion stream, hot spot).
Multiple variable contributions to He {\footnotesize{II}} emission 
are seen in other ULXs \citep{motch14}, and have prevented 
dynamical mass measurements \citep{roberts11}. 
Rapid, stochastic changes in the outflow 
density and temperature can easily change the X-ray luminosity 
(as shown by the rapid variability in the {\it Chandra} 
observations) and the irradiating flux, and hence change the relative contribution 
of different He {\footnotesize{II}} components. 
Another reason why we need to be cautious about the period measurement 
is that the sampling of radial velocities presented by \cite{liu13} is very sparse  
and does not even include a complete cycle. Two unrelated 
velocity dips around day 0 and day 90 (Fig.~2 in \citealt{liu13}), 
perhaps due to a contaminating component of He {\footnotesize{II}},     
could explain the data equally well. 
With the current sampling, it is impossible to rule out periods 
smaller than 3 days (which would allow the donor star 
to fill its Roche lobe) or longer than 100 days.

\subsection{Contribution to the optical/UV emission}

One of the main predictions of our model is that when a ULS 
is fainter in the soft X-ray band, it must be brighter in the UV. 
The dimmest observations in which a thermal component is significantly detected 
show a characteristic blackbody temperature $\approx$50 eV ($\approx$600,000 K) 
and a characteristic radius $\approx$100,000 km. This is the case not just 
for this particular ULS in M\,101, but also for most other ULSs 
\citep{urquhart15}. At even lower temperatures, ULSs 
are simply impossible to detect with {\it Chandra} or {\it XMM-Newton}; 
however, in those cases, can the optically thick photosphere be directly 
detected in the UV band, for example with the {\it Hubble Space 
Telescope} ({\it HST})? Assuming a reddening $E(B-V) = 0.025$ (roughly 
corresponding to the line-of-sight $N_{\rm H} \approx 1.5 \times 10^{20}$ 
cm$^{-2}$), $T_{\rm bb} = 50$ eV, $R_{\rm bb} = 10^5$ km at the distance 
of M\,101, the observed 
flux density at 2372.8 \AA\ (effective wavelength of the F225W filter 
on {\it HST}'s Wide Field Camera 3) is $F_{\nu} \approx 1.8 \times 10^{-8}$ Jy. 
This would result in a count rate of only $\approx$0.02 electrons s$^{-1}$; 
it would take an impossible 44,000 s to reach a signal-to-noise ratio of 7.
The same blackbody spectrum gives a flux density 
$F_{\nu} \approx 6 \times 10^{-9}$ Jy in the B band, corresponding 
to $B \approx 29.6$ mag (again, currently undetectable). Besides, such low 
optical fluxes are swamped by the optical emission from the outer 
part of the accretion disk and the donor star; in the case of the M\,101 ULS, 
the observed optical counterpart 
has a V-band flux density $\approx 1.5 \times 10^{-6}$ Jy 
($V \approx 23.5$ mag: \citealt{liu09}).

There is, however, a possibility that at some epochs the photosphere 
expands to even larger radii and the blackbody temperature becomes even 
cooler. At some point, the source may become detectable as an ultraluminous 
UV source. Such an object may have similarities with the compact, ultraluminous UV source \citep{kaaret10} detected in the core of the ionized nebula MF16 \citep{dunne00,blair01} in the galaxy NGC\,6946. In that case, the emission from the powerful BH may have three components \citep{roberts03,abolmasov08,kaaret10}: an ultraluminous X-ray component, a near-UV component with characteristic blackbody temperature $\approx$23,000--32,000 K (consistent with the outer accretion disk and/or a massive donor star), and a far-UV source with characteristic temperature $\approx$140,000 K and characteristic radius $\approx$$2 \times 10^6$ km $\approx$$3R_{\odot}$. The far-UV component is not directly imaged but is inferred from the ionizing flux required to produce the observed He {\footnotesize{II}} $\lambda$4686 line emission from the optical nebula \citep{abolmasov08}. There are not many astrophysical structures with the right size and temperature that can plausibly produce this far-UV component. It might be the optically thick photosphere of a large-scale outflow, even larger and cooler than what we modelled for the M\,101 ULS; however, direct two-component X-ray emission is visible in the MF16 ULX \citep{roberts03}. The two scenarios are not in contradiction: the ULX may have been engulfed by a thick outflow a few decades ago, when the ionizing far-UV photons were emitted, and may now be in a lower accretion regime in which we see the X-ray emission from the inner accretion flow. Scaled to the distance of M\,101, a hypothetical outflow photosphere with $T_{\rm bb} \approx 140,000$ K and $R_{\rm bb} \approx 2 \times 10^6$ km would have a flux density $F_{\nu} \approx 1.6 \times 10^{-6}$ Jy in the F225W filter of {\it HST}/WFC3 (corresponding to $\approx$2 electrons s$^{-1}$); it would be seen in the optical band at $B \approx 24.5$ mag, $V \approx 25$ mag.

\subsection{Absorption edges and harder emission tails}

Even at the modest spectral resolution provided by {\it Chandra}/ACIS, 
we have confirmed the presence of absorption edges 
at $E \approx 0.95$--1.05 keV, at some (but not all) epochs. 
The presence of (transient) edges was already noted as a ULS feature 
by \citet{kong04}; see also \citet{urquhart15}
for further examples of ULSs with edges. 
Such absorption edges are not seen in two-component ULXs, which may be 
further evidence that ULSs are seen through thicker wind. The presence 
of absorption edges at various energies (in particular from O{\small VIII}) 
was predicted (but not yet observationally tested) 
for SS\,433 as a key signature 
of super-critical outflows \citep{fabrika06,fabrika07}. 
 
At some epochs, we found significant residuals (particularly around 0.7--2 keV) 
in addition to the dominant, supersoft blackbody 
component. Such residuals can be described as a hard excess, 
or a hard emission tail, with a luminosity of a few $10^{38}$ erg 
s$^{-1}$ in the brightest epochs. 
We showed that multi-temperature, optically thin 
thermal plasma ({\it e.g.}, {\it mekal} in {\small {XSPEC}}) provides 
a good phenomenological fit to this harder emission.  
Similar spectral features around 1 keV, contributing a luminosity 
of a few $10^{38}$ erg s$^{-1}$, have been seen in some two-component 
ULXs ({\it e.g.}, NGC\,5408 X-1 and NGC\,6946 X-1) and have also been 
successfully fitted with thermal plasma emission 
\citep{middleton14,strohmayer07,stobbart06}.
However, it was also argued \citep{middleton14} that such features 
might instead be caused by broadened and blue-shifted absorption lines 
in a fast, partly ionized, optically-thin outflow.
We do not have enough spectral resolution and signal-to-noise ratio 
to test between those two scenarios for the M\,101 ULS. 
In either case, there is evidence both of an optically thick thermal 
emitter (which we argue is the photosphere of an optically thick outflow 
rather than the disk surface) and of a hotter component 
with absorption and/or emission features. Determining the origin 
of the hard tail and of the line features imprinted on it 
is beyond the scope of this work.

If all the excess hard emission comes from thermal-plasma components, 
we can use the model normalization to estimate 
the total amount of emitting gas visible to us. For ObsID 934-high, 
we have $\int n_e n_{\rm H} \, dV \approx 4 \times 10^{61}$ cm$^{-3}$. 
Taking the fitted value $R_{\rm bb} \approx 10^9$ cm as a characteristic 
length scale for the emitting gas, 
so that $V \approx 10^{27}$ cm$^3$, we obtain 
$n_e \approx 2 \times 10^{17}$ cm$^{-3}$,  
that is $\rho \approx 3 \times 10^{-7}$ g cm$^{-3}$ for ionized hydrogen gas.
We obtain a similar back-of-the-envelope estimate for the emission 
from ObsID 4737. For ObsID 934-medium, the estimated amount 
of optically-thin gas is slightly lower, $\rho \approx 10^{-7}$ g cm$^{-3}$.
This characteristic density compares well with 
the density range of the clumpy medium in the MHD simulations of 
\citet{takeuchi13} (see in particular their Fig.~1), 
at a comparable radial distance from the BH ($\sqrt{R^2 + z^2} \sim 10^4$ km).

Hard X-ray emission (modelled with bremsstrahlung or thermal plasma) 
from optically thin plasma,  
in addition to the dominant blackbody-like supersoft component, 
is often found in classical novae 
\citep{orio96,balman98,sokoloski06,hernanz10,li12}. 
It is generally explained with internal shocks in the expanding 
envelope, and/or shocks between the fast wind and circumstellar medium. 
One possible scenario applicable to the M\,101 ULS, in which the donor star is 
a Wolf-Rayet, is that some hard X-ray emission originates 
from interactions and shocks between the BH outflow and the Wolf-Rayet wind. 
This can perhaps explain the faint ($L_{0.3-10} \approx 1 \times 10^{37}$ 
erg s$^{-1}$) power-law component significantly detected 
only in the very deep, stacked spectrum of the dim-state observation 
but consistent with being present at all times. 
On the other hand, the much stronger hard X-ray component  
seen especially in ObsID 934 and ObsID 4737 
varies on the same short timescales as the soft-band emission. 
This is difficult to explain if one component comes from stellar wind 
interactions at $R \ga 10^{11}$ cm and the other from much  
nearer the BH ($R_{\rm bb} \approx 10^9$ cm). Therefore, we suggest 
that the hard tail is emitted either from the same region 
as the optically thick blackbody (perhaps a clumpy outflow, 
with optically thick clouds in between a hotter, lower-density medium), 
or even from smaller radii. 


\section{Conclusions}

The supersoft thermal spectrum of the M\,101 ULS (as well as 
those of a few other ULSs) has sometimes been interpreted 
as disk emission. Here, we re-examined the X-ray spectral 
and timing data from a series of {\it Chandra} and {\it XMM-Newton} 
observations, and discussed the main problems and internal 
inconsistencies of that interpretation. Instead, we showed 
that a model based on the photosphere of an optically thick outflow 
is consistent with the empirical data. For example, 
we showed that for $\dot{m} \approx 400$--700, our phenomenological model 
predicts (for BH masses $\approx 10 M_{\odot}$) 
blackbody temperatures $\sim$50--130 eV, characteristic radii 
$\sim$10,000--100,000 km, and bolometric luminosities 
$L^{\rm bol}_{\rm bb} \sim$ a few times $10^{39}$ erg s$^{-1}$, 
in agreement with the observations. 
In this scenario, the apparent brightness changes of the M\,101 ULS 
are mostly due to fast changes in the effective photospheric radius within 
a clumpy, fast outflow. When the photosphere expands, 
its characteristic temperature moves out of the {\it Chandra} 
band and into the far-UV band. The accretion rate is always highly 
super-critical, and the apparent faint states would appear 
just as luminous if observed in the far-UV.

Assuming that the massive wind is launched from near the spherization radius, 
and using a simple spherical and uniform analytic approximation for the outflow,
we have argued that there is a critical accretion rate 
($\dot{m} \sim$ a few 100, 
corresponding to $\dot{M} \sim 10^{-4} M_{\odot}$ yr$^{-1}$ for a stellar-mass BH) 
above which the outflow becomes effectively optically thick and completely 
shrouds the harder emission from the inner part of the accretion flow 
($R < R_{\rm sp}$).
In a more realistic, non-spherically-symmetric model, the wind is thicker 
when seen at higher inclination, and the corresponding threshold 
in $\dot{m}$ will have an angle dependence.
We suggested that if the accretion rate drops below this limit, 
and the photosphere shrinks, we may see a harder tail 
re-emerge in the observed spectrum, with a slope and high-energy break 
depending on the scattering optical depth in the wind \citep{sutton13}.

The soft thermal component of standard two-component ULXs 
is typically $\approx$130--300 eV \citep{miller04,kajava09,stobbart06}, 
and some authors have suggested \citep{soria07,kajava09} that 
the temperature decreases with increasing accretion rate 
(but see \citealt{miller13} for an opposite interpretation).
On the other hand, the typical blackbody temperature 
of ULSs is $\approx$50--150 eV. We suggest that 
$T_{\rm bb} \approx T_{\rm disk}(R = R_{\rm sp}) \approx 100$--150 eV 
is the critical threshold at which an outflow photosphere 
develops and shrouds the ULX: below those temperatures, 
hard energy tails are rarely seen, while they are usually 
observed in sources at higher temperatures. 
A comparison between the fitted temperature and luminosities 
of a larger sample of ULSs and ULXs is left to a companion paper 
\citep{urquhart15}.
In support of our suggested link between ULSs and ULXs, 
we noted that in the epochs when the fitted blackbody radius of the M\,101 ULS 
is larger, the temperature is lower 
and the spectrum is well modelled with a single blackbody component. 
In some of the epochs when the radius is smaller and the temperature higher, 
we found comparatively strong emission above 1.5 keV, which is consistent 
either with an additional, hotter thermal plasma component or with 
an inverse-Compton tail.
 
We also argued that the accretion rates required to produce a ULS  
in our scenario (especially if viewed at high inclination) 
are extremely high but not physically impossible: 
there is at least one source 
in our Galaxy (SS\,433) with a comparable accretion rate. 
Recent theoretical models of binary evolution \citep{wiktorowicz15} 
support the existence of such systems, provided that the donor star 
is filling its Roche lobe.
In the specific case of the M\,101 ULS, this requirement appears 
to be inconsistent with a claimed orbital period 
of $\approx$8 days \citep{liu13}. We argued that such optical 
variability measurement may not correspond to the true period, 
and further investigations on this issue are needed (but are beyond 
the scope of this paper).

\section*{Acknowledgments}

We thank Ryan Urquhart, Hua Feng, Jan-Uwe Ness, Manfred Pakull, 
Christian Motch, Doug Swartz, Chris Done, 
Yan-Fei Jiang, Rosanne Di Stefano, James Miller-Jones, Tom Russell, 
Peter Curran for useful suggestions and discussions. 
We also thank the anonymous referee for many useful suggestions.
RS acknowledges support from a Curtin University Senior Research Fellowship. 
He is also grateful for support and hospitality at Tsinghua University (Beijing) 
and Strasbourg Observatory during part of this work. 
This paper benefitted from discussions at the 2015 International Space Science Institute 
workshop ``The extreme physics of Eddington and super-Eddington accretion onto Black Holes''
in Bern, Switzerland (team PIs: Diego Altamirano \& Omer Blaes).

\appendix

\section[]{X-ray spectral parameters}

In the following tables, we list the best-fitting spectral parameters for a selection of spectral models and epochs. More specifically, we show (Table A1) how the high-count-rate interval of ObsID 934 requires harder components in addition to the soft thermal emission (this confirms the results of \citealt{mukai03} over the same time intervals). 
We then present (Table A2) a simultaneous fit of high-, intermediate- and low-count-rate intervals of ObsID 934, where we have fixed the intrinsic column density and the temperature of three thermal-plasma components, but we let their normalizations free; the temperature and normalization of the dominant blackbody component are also free. This model is illustrated in Figure 5. We then repeat the simultaneous fit to the same three intervals of ObsID 934 using two thermal-plasma components, accounting for the emission features at $\sim 0.5$--$1$ keV, and one Comptonization component, accounting for the hard excess above 1 keV (Table A3). We find that the model that includes the Comptonization component is statistically equivalent to the one that includes instead a high-temperature thermal plasma component. Finally, we list the best-fitting parameters for all other epochs: Table A4 is for the {\it Chandra} observations (plotted in Figures 3,4), and Table A5 for the {\it XMM-Newton} observation (Figure 8).  


\begin{table*}
\begin{center}
\begin{tabular}{lrrrrr}
\hline
 & Model 1 & Model 2 & Model 3  & Model 4 & Model 5  \\
\hline
Parameter &  \multicolumn{5}{c}{Value}  \\
\hline\\[-5pt]
$g_0$    &  & $[1]$ & $[1]$ & $[1]$ & $[1]$\\[5pt]
$\alpha$ 
           & & $0.36^{+0.07}_{-0.09}$ & $0.34^{+0.08}_{-0.08}$ & $0.25^{+0.08}_{-0.08}$  & $0.21^{+0.09}_{-0.07}$ \\[5pt]
psffrac &   & $[0.85]$ & $[0.85]$ & $[0.85]$ & $[0.85]$ \\[5pt]
$N_{\rm H,Gal}$ ($10^{20}$ cm$^{-2}$)  
         & $[1.5]$ & $[1.5]$ & $[1.5]$   & $[1.5]$ & $[1.5]$ \\[5pt]
$N_{\rm H,int}$ ($10^{20}$ cm$^{-2}$) 
         & $<0.8$ & $ 1.9^{+2.1}_{-1.8}$ & $ 2.7^{+2.1}_{-1.7}$ & $2.2^{+1.5}_{-0.8}$ & $5.6^{+1.4}_{-0.7}$ \\[5pt]
$kT_{\rm bb}$ (eV) 
         & $178^{+3}_{-3}$ & $159^{+9}_{-9}$ & $140^{+9}_{-11}$ & $138^{+6}_{-7}$ & \\[5pt]
$R_{\rm bb}$ ($10^3$ km) 
         & $4.3^{+0.3}_{-0.2}$ & $7.1^{+0.2}_{-0.9}$ & $9.2^{+3.0}_{-1.2}$ & $9.1^{+0.4}_{-0.4}$ & \\[5pt]
$kT_{\rm in}$ (eV) 
         &  & & & & $174^{+6}_{-7}$  \\[5pt]
$r_{\rm in} \sqrt{\cos \theta}$ ($10^3$ km) 
         &  & &  & & $5.9^{+0.3}_{-0.3}$ \\[5pt]
$kT_{\rm mekal1}$ (keV)
          & & & $0.88^{+0.13}_{-0.10}$ & $0.76^{+0.09}_{-0.09}$ & $0.72^{+0.09}_{-0.09}$ \\[5pt]
$N_{\rm mekal1}$ 
          & & & $4.8^{+1.5}_{-1.3} \times 10^{-5}$ & $4.1^{+0.8}_{-0.8} \times 10^{-5}$ & $4.8^{+0.8}_{-0.8} \times 10^{-5}$  \\[5pt]
$kT_{\rm mekal2}$ (keV)
          & & & & $1.6^{+1.3}_{-0.4}$ & $1.7^{+1.2}_{-0.4}$\\[5pt]
$N_{\rm mekal2}$ 
          & & & & $4.0^{+1.5}_{-1.5} \times 10^{-5}$ & $4.0^{+1.5}_{-1.5} \times 10^{-5}$\\[5pt]
\hline\\[-5pt]
$\chi^2_{\nu}$ & $2.23~(176.2/79)$ & $1.65~(128.7/78)$ & $1.23~(93.2/76)$ & $1.15~(84.9/74)$ & $1.16~(85.5/74)$\\[5pt]
\hline\\[-5pt]
$f_{0.3-10}$ ($10^{-13}$ erg cm$^{-2}$ s$^{-1}$) 
          & $3.8^{+0.2}_{-0.2}$ & $5.6^{+0.2}_{-0.2}$ & $5.7^{+0.2}_{-0.2}$ & $6.1^{+0.2}_{-0.2}$ & $6.1^{+0.2}_{-0.2}$  \\[5pt]
$L_{0.3-10}$ ($10^{39}$ erg s$^{-1}$) 
         & $2.0^{+0.1}_{-0.1}$ & $3.4^{+0.2}_{-0.2}$ & $3.8^{+0.2}_{-0.2}$ & $3.9^{+0.2}_{-0.2}$ & $5.2^{+0.3}_{-0.3}$   \\[5pt]
$L^{\rm bol}_{\rm bb}$ ($10^{39}$ erg s$^{-1}$) 
          & $2.4^{+0.1}_{-0.1}$ & $4.1^{+0.7}_{-0.5}$ & $4.1^{+0.7}_{-0.2}$ & $4.0^{+0.2}_{-0.1}$ &  \\[5pt]
$L^{\rm bol}_{\rm dbb} \cos \theta$ ($10^{39}$ erg s$^{-1}$) 
          &  &  &  &  & $4.1^{+0.2}_{-0.2}$ \\[5pt]
\hline
\end{tabular} 
\end{center}
\caption{Best-fitting spectral parameters for the high-count-rate intra-observation 
interval of {\it Chandra} ObsID 934, fitted with four different models. 
Model 1 is {\it tbabs} $\times$ {\it tbabs} $\times$ {\it blackbody}. 
Model 2 is {\it pileup} $\times$ {\it tbabs} $\times$ {\it tbabs} 
$\times$ {\it blackbody}. 
Model 3 is {\it pileup} $\times$ {\it tbabs} $\times$ {\it tbabs}  $\times$ 
({\it mekal} $+$ {\it mekal} $+$ {\it blackbody}).
Model 4 is {\it pileup} $\times$ {\it tbabs} $\times$ {\it tbabs}  $\times$ 
({\it mekal} $+$ {\it mekal} $+$ {\it diskbb}).
The {\it mekal} normalization is in units 
of $10^{-14}/(4\pi d^2)\, \int n_e n_{\rm H} \, dV$.
Errors indicate the 90\% confidence interval for each parameter of interest.
See also Figure 3.}
\label{taba1}
\end{table*}


\begin{table*}
\begin{center}
\begin{tabular}{lrrr}
\hline
 & Interval 1 (high) & Interval 2 (medium) & Interval 3 (low) \\
\hline
Parameter &  \multicolumn{3}{c}{Value}  \\
\hline\\[-5pt]
$g_0$    & $[1]$ & $[1]$ & $[1]$  \\[5pt]
$\alpha$ 
           & $0.27^{+0.08}_{-0.08}$ & $0.34^{+0.09}_{-0.10}$ & $[0]$  \\[5pt]
psffrac & $[0.85]$ & $[0.85]$ & $[0.85]$ \\[5pt]
$N_{\rm H,Gal}$ ($10^{20}$ cm$^{-2}$)  
         & $[1.5]$ & $[1.5]$ & $[1.5]$   \\[5pt]
$N_{\rm H,int}$ ($10^{20}$ cm$^{-2}$) 
         & $(4.1^{+0.4}_{-0.4})$   & $(4.1^{+0.4}_{-0.4})$  & $(4.1^{+0.4}_{-0.4})$ \\[5pt]
$kT_{\rm bb}$ (eV) 
         & $135^{+3}_{-4}$ & $119^{+3}_{-3}$ & $90^{+3}_{-3}$  \\[5pt]
$R_{\rm bb}$ ($10^3$ km) 
         & $10.2^{+0.4}_{-0.3}$ & $10.5^{+0.3}_{-0.3}$ & $18.4^{+0.9}_{-1.0}$ \\[5pt]
$kT_{\rm edge}$ (keV)
          & $(1.07^{+0.03}_{-0.05})$ & $(1.07^{+0.03}_{-0.05})$ & $(1.07^{+0.03}_{-0.05})$ \\[5pt]
$\tau_{\rm edge}$ 
          & $[0]$ & $2.1^{+1.3}_{-0.8}$ &  $[0]$\\[5pt]
$kT_{\rm mekal1}$ (keV)
          & $(0.61^{+0.06}_{-0.06})$ & $(0.61^{+0.06}_{-0.06})$ & $(0.61^{+0.06}_{-0.06})$ \\[5pt]
$N_{\rm mekal1}$ 
        & $1.7^{+1.0}_{-1.1} \times 10^{-5}$ & $2.3^{+0.5}_{-0.4} \times 10^{-5}$& $[0]$ \\[5pt]
$kT_{\rm mekal2}$ (keV)
          & $(0.98^{+0.16}_{-0.17})$ & $(0.98^{+0.16}_{-0.17})$ & $(0.98^{+0.16}_{-0.17})$\\[5pt]
$N_{\rm mekal2}$ 
          & $3.9^{+1.4}_{-1.4} \times 10^{-5}$ & $ <0.30 \times 10^{-5}$ & $[0]$ \\[5pt]
$kT_{\rm mekal3}$ (keV)
          & $(2.5^{+\ast}_{-1.2})$ & $(2.5^{+\ast}_{-1.2})$ & $(2.5^{+\ast}_{-1.2})$\\[5pt]
$N_{\rm mekal3}$ 
          & $1.9^{+1.9}_{-1.6} \times 10^{-5}$ & $ <0.56 \times 10^{-5}$ & $[0]$ \\[5pt]
\hline\\[-5pt]
$\chi^2_{\nu}$ & $(1.21~(174.6/144))$ & $(1.21~(174.6/144))$ & $(1.21~(174.6/144))$\\[5pt]
\hline\\[-5pt]
$f_{0.3-10}$ ($10^{-13}$ erg cm$^{-2}$ s$^{-1}$) 
          & $5.9^{+0.2}_{-0.2}$ & $2.9^{+0.2}_{-0.2}$ & $1.6^{+0.3}_{-0.3}$  \\[5pt]
$L_{0.3-10}$ ($10^{39}$ erg s$^{-1}$) 
         & $4.3^{+0.2}_{-0.2}$ & $2.3^{+0.2}_{-0.2}$ & $1.5^{+0.3}_{-0.3}$   \\[5pt]
$L^{\rm bol}_{\rm bb}$ ($10^{39}$ erg s$^{-1}$) 
          & $4.6^{+0.2}_{-0.3}$ & $2.9^{+0.2}_{-0.2}$ & $2.9^{+0.3}_{-0.3}$  \\[5pt]
\hline
\end{tabular} 
\end{center}
\caption{Best-fitting spectral parameters for {\it Chandra} ObsID 934, 
fitted with {\it pileup} $\times$ {\it tbabs} $\times$ {\it tbabs} $\times$ {\it edge}
$\times$ ({\it mekal} $+$ {\it mekal} $+$ {\it mekal} $+$ {\it blackbody}). 
This was a simultaneous fit to the spectra in the three intra-observation 
intervals, defined from the observed count rates (high, medium, low).  
Parameters listed in square brackets were frozen during the fit;
parameters listed in round brackets were free but locked for all three spectra; 
all other parameters were left free to vary independently.
The {\it mekal} normalization is in units 
of $10^{-14}/(4\pi d^2)\, \int n_e n_{\rm H} \, dV$.
Errors indicate the 90\% confidence interval for each parameter of interest. 
The fitted spectra and $\chi^2$ residuals are plotted in Figure 2 (top left panel).}
\label{taba2}
\end{table*}

\begin{table*}
\begin{center}
\begin{tabular}{lrrr}
\hline
 & Interval 1 (high) & Interval 2 (medium) & Interval 3 (low) \\
\hline
Parameter &  \multicolumn{3}{c}{Value}  \\
\hline\\[-5pt]
$g_0$    & $[1]$ & $[1]$ & $[1]$  \\[5pt]
$\alpha$ 
           & $0.27^{+0.09}_{-0.08}$ & $0.34^{+0.09}_{-0.10}$ & $[0]$  \\[5pt]
psffrac & $[0.85]$ & $[0.85]$ & $[0.85]$ \\[5pt]
$N_{\rm H,Gal}$ ($10^{20}$ cm$^{-2}$)  
         & $[1.5]$ & $[1.5]$ & $[1.5]$   \\[5pt]
$N_{\rm H,int}$ ($10^{20}$ cm$^{-2}$) 
         & $(4.2^{+0.4}_{-0.4})$   & $(4.2^{+0.4}_{-0.4})$  & $(4.2^{+0.4}_{-0.4})$ \\[5pt]
$kT_{\rm bb}$ (eV) 
         & $134^{+2}_{-4}$ & $119^{+2}_{-2}$ & $90^{+3}_{-3}$  \\[5pt]
$R_{\rm bb}$ ($10^3$ km) 
         & $10.5^{+0.3}_{-0.3}$ & $10.5^{+0.3}_{-0.3}$ & $18.4^{+0.9}_{-1.0}$ \\[5pt]
$kT_{\rm edge}$ (keV)
          & $(1.07^{+0.03}_{-0.05})$ & $(1.07^{+0.03}_{-0.05})$ & $(1.07^{+0.03}_{-0.05})$ \\[5pt]
$\tau_{\rm edge}$ 
          & $[0]$ & $2.1^{+1.3}_{-0.8}$ &  $[0]$\\[5pt]
$kT_{\rm mekal1}$ (keV)
          & $(0.61^{+0.06}_{-0.07})$ & $(0.61^{+0.06}_{-0.07})$ & $(0.61^{+0.06}_{-0.07})$ \\[5pt]
$N_{\rm mekal1}$ 
        & $1.9^{+1.2}_{-1.2} \times 10^{-5}$ & $2.3^{+0.4}_{-0.4} \times 10^{-5}$& $[0]$ \\[5pt]
$kT_{\rm mekal2}$ (keV)
          & $(0.99^{+0.16}_{-0.15})$ & $(0.99^{+0.16}_{-0.15})$ & $(0.99^{+0.16}_{-0.15})$\\[5pt]
$N_{\rm mekal2}$ 
          & $4.0^{+1.4}_{-1.2} \times 10^{-5}$ & $ <0.30 \times 10^{-5}$ & $[0]$ \\[5pt]
$kT_0$ (eV)
          & $(134^{+2}_{-4})$ & $(119^{+2}_{-2})$ & $(90^{+3}_{-3})$\\[5pt]
$kT_e$ (keV)
          & $[4.0]$ & $[4.0]$ & $[4.0]$\\[5pt]
$\tau_e$ (keV)
          & $(>2.0)$ & $(>2.0)$ & $(>2.0)$\\[5pt]
$N_{\rm comptt}$ 
          & $7.2^{+7.3}_{-6.6} \times 10^{-6}$ & $ <2.6 \times 10^{-6}$ & $[0]$ \\[5pt]
\hline\\[-5pt]
$\chi^2_{\nu}$ & $(1.21~(174.1/144))$ & $(1.21~(174.1/144))$ & $(1.21~(174.1/144))$\\[5pt]
\hline\\[-5pt]
$f_{0.3-10}$ ($10^{-13}$ erg cm$^{-2}$ s$^{-1}$) 
          & $5.9^{+0.2}_{-0.2}$ & $2.9^{+0.2}_{-0.2}$ & $1.6^{+0.3}_{-0.3}$  \\[5pt]
$L_{0.3-10}$ ($10^{39}$ erg s$^{-1}$) 
         & $4.4^{+0.2}_{-0.2}$ & $2.3^{+0.2}_{-0.2}$ & $1.5^{+0.3}_{-0.3}$   \\[5pt]
$L^{\rm bol}_{\rm bb}$ ($10^{39}$ erg s$^{-1}$) 
          & $4.5^{+0.2}_{-0.3}$ & $2.9^{+0.2}_{-0.2}$ & $2.9^{+0.3}_{-0.3}$  \\[5pt]
\hline
\end{tabular} 
\end{center}
\caption{Best-fitting spectral parameters for the three intra-observation 
intervals of ObsID 934, 
fitted with {\it pileup} $\times$ {\it tbabs} $\times$ {\it tbabs} $\times$ {\it edge}
$\times$ ({\it mekal} $+$ {\it mekal} $+$ {\it blackbody} $+$ {\it comptt}). 
Parameters listed in square brackets were frozen during the fit;
parameters listed in round brackets were free but locked for all three spectra; 
all other parameters were left free to vary independently.
The {\it mekal} normalization is in units 
of $10^{-14}/(4\pi d^2)\, \int n_e n_{\rm H} \, dV$.
Errors indicate the 90\% confidence interval for each parameter of interest. 
This model is statistically and visually indistinguishable from the three-temperature 
{\it mekal} model of Table A2.}
\label{taba3}
\end{table*}

\begin{table*}
\begin{center}
\begin{tabular}{lrrrrrr}
\hline
 & 2000 Oct 29 & 2004 Jul 5--11 & 2004 Dec 22--24 & 2004 Dec 30 & 2005 Jan 1 & Faint state\\
\hline
Parameter &  \multicolumn{6}{c}{Value}  \\
\hline\\[-5pt]
$g_0$    &  &  & & & $[1]$ & \\[5pt]
$\alpha$ 
           & &  & & & $1.00^{+\ast}_{-0.25}$ &  \\[5pt]
psffrac &   &  & & & $[0.85]$ &  \\[5pt]
$N_{\rm H,Gal}$ ($10^{20}$ cm$^{-2}$)  
         & $[1.5]$ & $[1.5]$  & $[1.5]$ & $[1.5]$ & $[1.5]$  & $[1.5]$ \\[5pt]
$N_{\rm H,int}$ ($10^{20}$ cm$^{-2}$) 
         & $7.5^{+7.0}_{-4.8}$ & $9.7^{+4.0}_{-3.7}$ & $20.4^{+13.9}_{-11.5}$   & $12.2^{+6.1}_{-5.4}$  & $13.4^{+3.2}_{-3.4}$&  $19.3^{+15.6}_{-13.3}$\\[5pt]
$kT_{\rm bb}$ (eV) 
         & $77^{+11}_{-10}$ & $69^{+7}_{-7}$  & $39^{+10}_{-\ast}$ & $75^{+6}_{-6}$  & $100^{+13}_{-10}$ & $53^{+8}_{-11}$ \\[5pt]
$R_{\rm bb}$ ($10^3$ km) 
         & $29.0^{+32.7}_{-14.4}$ & $47.0^{+32.6}_{-17.0}$ & $> 54$ & $43.5^{+29.3}_{-17.2}$ & $22.5^{+10.3}_{-4.7}$& $24.3^{+102.1}_{-10.8}$ \\[5pt]
$\Gamma$ (keV)
          & &  & & & & $1.9^{+0.9}_{-0.7}$\\[5pt]
$N_{\rm po}$ ($10^{-7}$)
          &  &  & & & & $3.8^{+2.7}_{-1.6}$\\[5pt]
$kT_{\rm edge}$ (keV)
          & &  $0.93^{+0.05}_{-0.04}$ & & & & \\[5pt]
$\tau_{\rm edge}$ 
          &  &  $2.1^{+1.6}_{-0.9}$ & & & & \\[5pt]
$kT_{\rm mekal1}$ (keV)
          & &$0.59^{+0.21}_{-0.26}$ & &  & $0.70^{+0.17}_{-0.13}$ &  \\[5pt]
$N_{\rm mekal1}$ 
          & &$2.9^{+4.6}_{-1.5} \times 10^{-6}$ & & & $3.1^{+1.5}_{-1.5} \times 10^{-5}$ &  \\[5pt]
$kT_{\rm mekal2}$ (keV)
          & & & & & $1.3^{+0.2}_{-0.2}$ & \\[5pt]
$N_{\rm mekal2}$ 
          & & & & & $4.3^{+1.7}_{-1.6} \times 10^{-5}$ & \\[5pt]
\hline\\[-5pt]
$\chi^2_{\nu}$ & $1.04~(12.4/12)$ & $1.17~(42.1/36)$ & & $0.87~(22.5/26)$ & $1.02~(63.0/62)$& $0.98~(9.8/10)$ \\[5pt]
Cash-stat &  &  & $26.3/44$ & & & \\[5pt]
\hline\\[-5pt]
$f_{0.3-10}$ ($10^{-13}$ erg cm$^{-2}$ s$^{-1}$) 
          & $1.1^{+0.1}_{-0.2}$ & $1.1^{+0.1}_{-0.1}$ & $\approx 0.1$ & $1.3^{+0.1}_{-0.1}$  & $3.2^{+0.1}_{-0.1}$&  $0.028^{+0.010}_{-0.007}$\\[5pt]
$L_{0.3-10}$ ($10^{39}$ erg s$^{-1}$) 
         & $1.6^{+1.8}_{-0.8}$ & $2.3^{+1.2}_{-0.8}$ & $\approx 1.5$ & $3.1^{+2.7}_{-1.3}$  & $5.5^{+1.3}_{-1.2}$&   $0.11^{+0.06}_{-0.04}$ \\[5pt]
$L^{\rm bol}_{\rm bb}$ ($10^{39}$ erg s$^{-1}$) 
          & $3.6^{+6.8}_{-2.0}$ & $6.6^{+5.9}_{-2.9}$ & $>2.1$ & $7.7^{+8.6}_{-3.8}$  & $6.6^{+1.2}_{-1.0}$ & $0.6^{+2.7}_{-0.2}$ \\[5pt]
\hline
\end{tabular} 
\end{center}
\caption{Best-fitting spectral parameters for the other six spectra 
(some of them single observations, some stacked) plotted in Figure 2.
For all spectra, we started with a simple {\it tbabs} $\times$ {\it tbabs} 
$\times$ {\it blackbody} model; we included a correction for pileup for 
the 2005 January 1 observation because of its high count rate. 
We then added an {\it edge} and/or {\it mekal} components for a few epochs,   
if they significantly improved the fit.
The {\it mekal} normalization is in units 
of $10^{-14}/(4\pi d^2)\, \int n_e n_{\rm H} \, dV$. The {\it powerlaw} 
normalization is in photons keV$^{-1}$ cm$^{-2}$ s$^{-1}$ at 1 keV.
Errors indicate the 90\% confidence interval for each parameter of interest.}
\label{taba4}
\end{table*}

\begin{table*}
\begin{center}
\begin{tabular}{lrr}
\hline
 & 2004 Jul 23 & 2005 Jan 8\\
\hline
Parameter &  \multicolumn{2}{c}{Value}  \\
\hline\\[-5pt]

$N_{\rm H,Gal}$ ($10^{20}$ cm$^{-2}$)  
         & $[1.5]$ & $[1.5]$ \\[5pt]
$N_{\rm H,int}$ ($10^{20}$ cm$^{-2}$) 
         & $9.8^{+15.4}_{-9.8}$ & $11.8^{+4.7}_{-4.2}$\\[5pt]
$kT_{\rm bb}$ (eV) 
         & $48^{+20}_{-25}$ & $56^{+5}_{-5}$   \\[5pt]
$R_{\rm bb}$ ($10^3$ km) 
         & $64.7^{+\ast}_{-54.1}$ & $101.5^{+82.7}_{-43.1}$ \\[5pt]
\hline\\[-5pt]
$\chi^2_{\nu}$  & $0.92~(7.4/8)$ & $0.80~(27.2/34)$ \\[5pt]
\hline\\[-5pt]
$f_{0.3-10}$ ($10^{-13}$ erg cm$^{-2}$ s$^{-1}$) 
          & $0.11^{+0.02}_{-0.02}$ & $0.8^{+0.1}_{-0.1}$ \\[5pt]
$L_{0.3-10}$ ($10^{39}$ erg s$^{-1}$) 
         & $0.39^{+\ast}_{-0.32}$  & $2.6^{+1.9}_{-1.0}$  \\[5pt]
$L^{\rm bol}_{\rm bb}$ ($10^{39}$ erg s$^{-1}$) 
          & $3.5^{+\ast}_{-3.3}$ & $12.8^{+17.9}_{-6.9}$ \\[5pt]
\hline
\end{tabular} 
\end{center}
\caption{Best-fitting spectral parameters for the combined 
{\it XMM-Newton}/EPIC spectrum of the 2004 July 23 and 2005 January 8 
observations, plotted in Figure 8.
Errors indicate the 90\% confidence interval for each parameter of interest.}
\label{taba5}
\end{table*}

\end{document}